\documentclass[twocolumn]{aastex63}

\newcommand{\avg}[1]{\left<#1\right>}
\newcommand{\yFeMgas}{y_{\text{Fe}}f_{\text{retained}}/M_{\text{gas}}}
\newcommand{\fret}{f_{\text{retained}}}
\newcommand{\NSN}{N_{\text{SN}}}
\newcommand{\fr}{f_r}
\newcommand{\fzero}{f_0}
\newcommand{\Mrmin}{M_{r,\text{min}}}
\newcommand{\MFe}{M_{\text{Fe}}}
\newcommand{\MEu}{M_{\text{Eu}}}
\newcommand{\MH}{M_{\text{H}}}
\newcommand{\XEu}{X_{\text{Eu}}}
\newcommand{\IQREu}{\text{IQR}_{\text{Eu}}}

\defcitealias{Roederer14b}{R14}

\received{xxx x, 2020}
\revised{xxx x, 2020}
\accepted{\today}
\submitjournal{AJ}

\shorttitle{R-Process Collapsars}
\shortauthors{Brauer et al.}
\graphicspath{{./}{figures/}}

\begin{document}

\title{Collapsar R-Process Yields Can Reproduce [Eu/Fe] Abundance Scatter in Metal-Poor Stars}

\correspondingauthor{Kaley Brauer}
\email{kbrauer@mit.edu}

\author[0000-0002-8810-858X]{Kaley Brauer}
\affiliation{Department of Physics and Kavli Institute for Astrophysics and Space Research, Massachusetts Institute of Technology, Cambridge, MA 02139, USA}
\affiliation{Joint Institute for Nuclear Astrophysics -- Center for Evolution of the Elements, USA}

\author[0000-0002-4863-8842]{Alexander~P.~Ji}
\affiliation{Observatories of the Carnegie Institution for Science, 813 Santa Barbara St., Pasadena, CA 91101, USA}
\affiliation{Joint Institute for Nuclear Astrophysics -- Center for Evolution of the Elements, USA}
\affiliation{Hubble Fellow}

\author[0000-0001-7081-0082]{Maria R. Drout}
\affiliation{David A. Dunlap Department of Astronomy and Astrophysics, University of Toronto \\ 50 St. George Street, Toronto, Ontario, M5S 3H4 Canada}
\affiliation{Observatories of the Carnegie Institution for Science, 813 Santa Barbara St., Pasadena, CA 91101, USA}

\author[0000-0002-2139-7145]{Anna Frebel}
\affiliation{Department of Physics and Kavli Institute for Astrophysics and Space Research, Massachusetts Institute of Technology, Cambridge, MA 02139, USA}
\affiliation{Joint Institute for Nuclear Astrophysics -- Center for Evolution of the Elements, USA}



\begin{abstract}

It is unclear if neutron star mergers can explain the observed $r$-process abundances of metal-poor stars. Collapsars, defined here as rotating massive stars whose collapse results in a rapidly accreting disk around a black hole that can launch jets, are a promising alternative. We find that we can produce a self-consistent model in which a population of collapsars with stochastic europium yields synthesizes all of the $r$-process material in metal-poor ([Fe/H] $<-2.5$) stars. Our model reproduces the observed scatter and evolution of scatter of [Eu/Fe] abundances. We find that if collapsars are the dominant $r$-process site for metal-poor stars, $r$-process synthesis may be linked to supernovae that produce long $\gamma$-ray bursts. Our results also allow for the possibility that core-collapse supernovae beyond those that launch $\gamma$-ray bursts also produce $r$-process material (e.g., potentially a subset of Type Ic-BL supernovae). Furthermore, we identify collapsar jet properties (isotropic energy, engine luminosity, or engine time) which may trace $r$-process yield and verify that the amount of $r$-process yield produced per collapsar in our model ($\sim 0.07 M_\odot$) is consistent with other independent estimates. In the future, achieving 0.05 dex precision on distribution scatter or a reliable selection function would further constrain our probe of $r$-process production. Our model would also hold for another prompt $r$-process site with a power-law yield, and work is needed to determine if, for example, fast-merging neutron stars can also explain abundance scatter.

\end{abstract}



\section{Introduction} \label{sec:intro}

Around half of the abundances of the heaviest isotopes in the periodic table, including gold and europium, are produced through the rapid neutron-capture process \citep[$r$-process,][]{Burbidge57,Cameron57}. Since the first discussion of the $r$-process in the 1950s, there has been debate over which astrophysical sites produce $r$-process material. 
Recently, the detection of an optical transient associated with the neutron star merger GW170817 \citep{LIGOGW170817a,coulter17} provided strong evidence for $r$-process production in neutron star mergers \citep[e.g.,][]{Drout17, Pian17}. Neutron star mergers thus appear to be a source of $r$-process elements, but it is unclear if they are the dominant source in the early universe. One concern stems from observations of $r$-process abundances in metal-poor ([Fe/H] $< -2.5$) stars in the Galactic halo.

It is unclear whether the delay time to form and coalesce a binary neutron star system is too long to provide $r$-process material to near-pristine gas before the formation of metal-poor stars \citep[e.g.,][]{Argast04,Skuladottir19,Cescutti15,Wehmeyer15,Haynes19,Kobayashi20}. Possible solutions include processes like inhomogeneous metal mixing or inefficient star formation mitigating the delay time \citep[e.g.,][]{Ishimaru15,Shen15,vandeVoort15,RamirezRuiz15,Ji16b,Dvorkin20} or common envelope producing a large number of rapidly merging neutron star binaries \citep[e.g.,][]{Beniamini16,Safarzadeh19,Zevin19,Andrews20}, but concerns have not been eradicated.

Natal kicks received from the supernova explosions that give birth to neutron stars may have also made it unlikely for small, early galaxies to retain neutron star binaries \citep{Bramante16,Beniamini16,Bonetti19}. For example, the highly $r$-process-enriched metal-poor stars in the ultra-faint dwarf galaxy Reticulum II could potentially be explained by a neutron star merger \citep{Ji16b}, but the natal kick would have to have been very small ($v < v_{esc} \sim 10-20 $ km s\textsuperscript{-1}) and/or the merger time extremely short to avoid kicking the binary out of the tiny galaxy \citep{Tarumi20,Safarzadeh19b,Bramante16}. This is in contrast to larger estimates of $20-140$ km s$^{-1}$ based on the offset distribution of short-duration $\gamma$-ray bursts from their host galaxies \citep{Fong2013}, and $5-5450$ km s$^{-1}$ from galactic double neutron star systems \citep{Wong2010}.

In light of these concerns --- coupled with the inference that the ejecta from GW170817 was dominated by an accretion disk wind, rather than dynamical tidal tails \citep[e.g.][]{siegel2019b} --- \citet{siegel19} revived the idea that collapsars (the supernova- and $\gamma$-ray-burst-triggering collapse of rapidly rotating massive stars) may be an important source of $r$-process material (see also, e.g., \citealt{MacFadyen99,McLaughlin2005}). In particular, the accretion disks formed in collapsars can have similar conditions to the $r$-process producing disk of GW170817. \citet{siegel19} found that for accretion rates $\gtrsim$ 10$^{-3}$ M$_\odot$ s$^{-1}$, these disks produce neutron-rich outflows that synthesize heavy $r$-process nuclei. They also found that collapsars can yield sufficient $r$-process material to explain over 80\% of the $r$-process content of the Universe. 
Although the electron fraction in collapsar disk winds is still debated \citep{Surman06,Miller19}, this is currently one of the most promising ways for core-collapse supernovae (CCSN) to make $r$-process elements (other than magnetorotationally driven CCSN, e.g., \citealt{Nishimura15}, but see \citealt{Mosta18}).

The scatter in the abundances of metal-poor stars is a useful probe of different $r$-process origins.
In this paper, we investigate collapsars as a source of the $r$-process in the early universe by investigating whether they can self-consistently reproduce the scatter of europium (Eu, Z = 63) in the most metal-poor stars. Our results also hold for any prompt $r$-process site with a power law distribution of effective $r$-process yields. As our representation was directly inspired by collapsar properties for the purposes of studying $r$-process collapsars, though, we call the $r$-process site in our model ``collapsars" and discuss alternative interpretations in more detail in Section \ref{subsec:NSMvscollapsar}.

Our model assumes the $r$-process material in metal-poor stars was formed exclusively in collapsars with stochastic $r$-process yields. Previous stochastic models primarily assume the $r$-process is produced in fixed amounts, but comes from multiple different sources and/or mixes into different environments \citep[e.g.,][]{Tsujimoto+Shigeyama14,Cescutti15,Wehmeyer15,Shen15}. 
In contrast, our model assumes the $r$-process source has an intrinsically stochastic production: each collapsar synthesizes a different amount of $r$-process material.

In Section \ref{sec:model}, we outline our stochastic collapsar enrichment model in which we assume each collapsar contributes an $r$-process yield that is independently drawn from a power law distribution, inspired by models of collapsar jet fits to $\gamma$-ray burst data. Our model is constrained using stellar abundance data described in Section \ref{sec:data}, and the parameter constraints are described in Section \ref{sec:results}. The implications of these results are discussed in Section \ref{sec:disc}, where we put our results in context with collapsar jet property distributions, different types of core-collapse supernovae, and different estimates for the amounts of $r$-process material which may be produced by collapsars. Our conclusions are summarized in Section \ref{sec:conc}.

\section{Collapsar $r$-Process Yield Model} \label{sec:model}

\begin{figure*}[t!]
\gridline{
	\fig{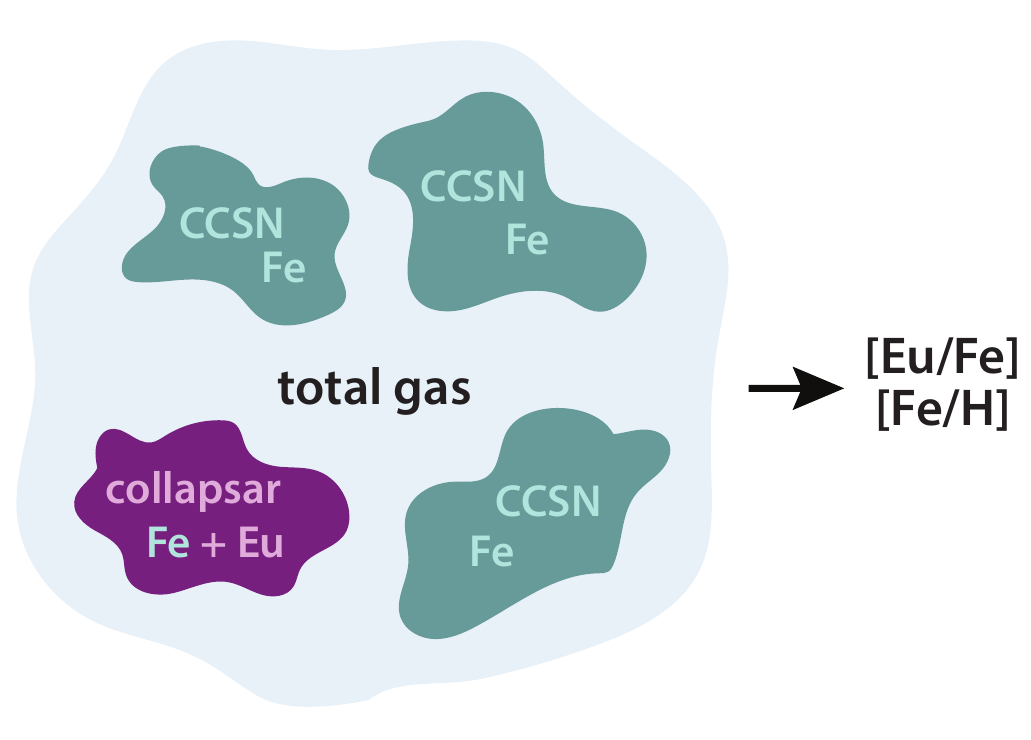}{0.45\textwidth}{(a) Schematic of our model. Core-collapse supernovae explode into the total gas, yielding iron (Fe), and some fraction of these are collapsars which also yield stochastic amounts of europium (Eu). This produces stars with different [Eu/Fe] and [Fe/H] abundances.}
	\fig{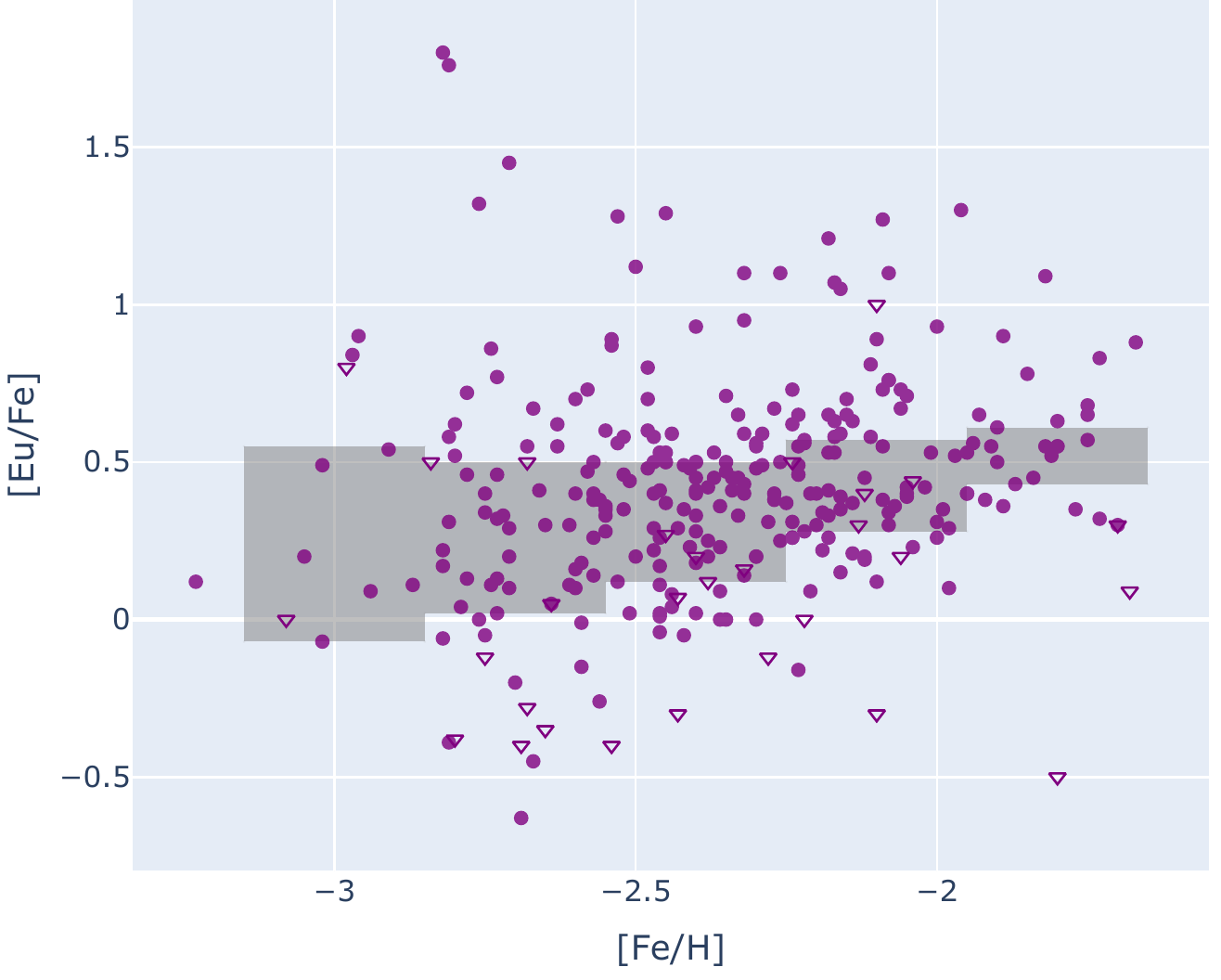}{0.4\textwidth}{(b) The $r$-process abundance scatter in the RPA data. The data is shown as dots (Eu detections) and triangles (Eu upper limits). The grey boxes show the estimated $\IQREu$ (the abundance scatter) in several metallicity bins, which decreases with increasing metallicity.}}
\caption{Schematic of our theoretical model and scatter plot of the stellar data our models attempts to reproduce. Our model attempts to reproduce the observed Eu scatter at low metallicity by assuming all Eu is produced by collapsars, which are a fraction of all core-collapse supernovae.
\label{fig:stellarscatter}}
\end{figure*}

The purpose of our model is to determine the distribution of $r$-process abundances (as measured by [Eu/Fe]) in a fixed metallicity bin (as measured by [Fe/H]). A schematic of the model can be seen in Figure \ref{fig:stellarscatter}a. The novel feature of our model is that we explicitly study whether variable $r$-process yields from a single class of r-process events could produce observed abundance scatter. This is in contrast to previous models \citep[e.g.,][]{Cescutti15,Ojima18,Shen15,vandeVoort15} which generally had a fixed yield per event and produced scatter through different $r$-process sites and/or different galactic environments.

\subsection{Defining ``Collapsar''} \label{subsec:definecollap}

The term ``collapsar'' typically refers to the collapse of a massive, rapidly-rotating star in which accretion onto a central black hole can produce a beamed jet, commonly evoked as the progenitors of long-duration $\gamma$-ray bursts (LGRBs). In the model described below, we more broadly use the term to encompass a population of core-collapse supernovae that produce heavy r-process material with a power-law distribution of yields. 
This definition is motivated by the traditional collapsar picture in which rapid accretion onto a compact object launches a collimated outflow wherein both the duration and luminosities of LGRBs are well-described by power laws \citep{petropoulou17,sobacchi17}. Connecting jet properties to $r$-process production is inspired by the possible connection between the accretion phase during which $r$-process material is produced (due to a sufficiently high accretion rate that neutronizes the disk) and the phase during which the collapsar jet is launched. In particular, \citet{siegel19} finds that the production of heavy r-process material requires $\dot{M} \gtrsim$ 10$^{-3}$ M$_\odot$ s$^{-1}$, closely matched to the accretion rates required for jet production \citep{MacFadyen99}. The results described below also hold for any prompt $r$-process site (e.g., occurring roughly concurrently with CCSN; this could potentially include fast-merging neutron stars) that lead to a power law distribution of heavy $r$-process material. We discuss other interpretations in Section \ref{subsec:NSMvscollapsar}.

In addition, our model does not require that a jet successfully breaks out of the progenitor star. While the most extreme $r$-process producing events require large fallback accretion disks and are likely associated with LGRBs, our model also includes events that eject smaller amounts of $r$-process material. Such systems may produce weaker outflows/jets and be observed as low-luminosity GRBs \citep[ll-GRBs; e.g.,][]{Bromberg11,petropoulou17}, relativistic supernovae \citep[e.g.,][]{Soderberg2010,Margutti2014}, or broad-lined Type Ic supernovae \citep[Type Ic-BL; e.g.,][]{Milisavljevic2015,Modjaz2016}.

\subsection{Basic Physical Set-up and Model Parameters}

We analytically model the abundance distribution as arising from a burst of core-collapse supernova enrichment, some fraction of which are collapsars that produce non-zero $r$-process yields.
Each core-collapse supernova produces an iron yield and each collapsar also produces an $r$-process yield that is independently and identically drawn from a power law distribution. We assume the typical star formed in these galaxies forms after the metal yields from all of the supernovae fall into and mix within the hydrogen gas of the system. During this process, some fraction of the metals are permanently lost from the galaxy due to gas outflows.

This model has five free parameters: 
\begin{enumerate}
\item $\NSN$: the number of core-collapse supernovae enriching the gas.
\item $\avg{N_r}$: the average number of collapsars (i.e. supernovae that produce non-zero amounts of $r$-process material).
\item $\Mrmin$: the minimum mass of $r$-process material that can be produced by a collapsar. 
\item $\alpha$: the power law exponent for our power law distribution of $r$-process yield produced per collapsar.
\item $y_{\rm{Fe,eff}}$: the effective iron yield per supernovae per unit gas mass. $y_{\rm{Fe,eff}} = \yFeMgas$, where $y_{\text{Fe}}$ is the iron yield per supernova, $\fret$ is the fraction of iron retained in the galaxy and not carried out of the system by gas outflows, and $M_{gas}$ is the total gas mass in the system.
\end{enumerate}

These parameters combine to yield the mean gas metallicity, the the fraction of supernovae that are collapsars ($f_r$), and the average yield per collapsar ($\avg{M_r}$), as described below.

\subsection{Gas Enrichment}

We determine the distribution of europium abundances produced by this model at a given metallicity. The mean metallicity of our stars is found from $\MFe/\MH = \NSN y_{\rm{Fe,eff}}$. The mass of iron and hydrogen is converted to [Fe/H] using a mean molecular weight of $\mu_{\text{Fe}}=56$ for Fe, $\log \epsilon_\odot(\text{Fe}) = 7.50$, and $\log \epsilon_\odot(\text{H}) = 12.00$ \citep{Asplund09}, where we use the stellar spectroscopist notation $\mbox{[X/Y]} \equiv \log N_{\text{X}} / N_{\text{Y}} - \log \left(N_{\text{X}}/N_{\text{Y}}\right)_\odot = \log(\frac{M_{\text{X}}\mu_{\text{Y}}}{M_{\text{Y}}\mu_{\text{X}}}) - (\log\epsilon_\odot(\text{X}) - \log\epsilon_\odot(\text{Y}))$. 

In order to determine the distribution of europium values, we enrich this gas with $N_r$ collapsars, which we draw stochastically from a Poisson distribution with mean $\avg{N_r} = \fr \NSN$.
Each collapsar contributes an $r$-process yield $M_r$ that is independently drawn from a power law distribution.

\begin{equation}
  p(M_r) \propto \left(\frac{M_r}{\Mrmin}\right)^{-\alpha} \quad M_r \geq \Mrmin  \label{eq:Mr}
\end{equation}

The $r$-process yield for a single explosion, $M_r$, can be converted to $\MEu$ by using the solar $r$-process mass fraction of europium compared to all nuclei with mass number A $> 70$. The mass fraction, $\XEu$, is approximately $10^{-3}$ ($1.75 \times 10^{-3}$, \citealt{Arnould07}; $9.77 \times 10^{-4}$, \citealt{Sneden08}). When converting total europium mass to [Eu/Fe], we use a mean molecular weight of $\mu_{\text{Fe}}=152$ for Eu and $\log \epsilon_\odot(\text{Eu}) = 0.52$. We also assume that europium and iron have the same retention fraction, $f_{\text{retained}}$, meaning the same fraction of both is lost from the galaxy.

Note that if $\alpha \leq 2$, then the average yield produced per collapsar diverges and our model would also require an upper cutoff to the amount of r-process material that can be produced by a single collapsar, $M_{r,\text{max}}$. However, when we compare to observed data in Section \ref{sec:results}, it will turn out our results imply $\alpha > 2$, in which case the average yield per collapsar is: 

\begin{equation}
\avg{M_r} = \Mrmin \frac{\alpha-1}{\alpha-2}
\end{equation}

While in principle $y_{\text{Fe}}$ is also stochastic, for simplicity we hold it constant. This is fine as long as $\fr$ is small, since variations in the Fe yield will average out.

Operationally, we create a model [Eu/Fe] distribution by considering several thousand instances of supernova enrichment. Each instance is a single data point in our modeled cumulative distribution function. For each instance, we draw an $N_r$ value and then draw $M_r$ for each of the $N_r$ collapsars. The total europium and iron masses retained in the galaxy in each instance are transformed into a [Eu/Fe] measurement. We also add a 0.1 dex Gaussian uncertainty to mimic observational errors.

\subsection{Constraining Model Parameters: Literature Estimates for Effective Iron Yields} \label{subsec:res_yfM}

The effective iron yield of core-collapse supernova per unit gas mass cannot be directly constrained from a sample of stellar abundance data. We constrain its value by combining estimates for each component parameter (recall $y_{\rm{Fe,eff}} = \yFeMgas$) from the literature. 

The fraction of retained metals is set to $\fret = 10^{-2 \pm 0.5}$,  assuming that metal-poor stars form early in small galaxies. Observationally, individual faint galaxies have $\fret$ in this range: the Milky Way's moderately faint dSphs (e.g., Ursa Minor) have kept less than 1\% of their metals \citep{Kirby11outflow}; while the faint but still star-forming galaxy Leo P has kept about 5\% of its metals \citep{McQuinn15}.
Theoretically, retaining about 1\% of metals in small galaxies reproduces the slope and normalization of the mass-metallicity relation \citep[e.g.,][]{Dekel03,Robertson05}. The retention fraction is also borne out in hydrodynamic galaxy simulations \citep[e.g.,][]{Emerick18}. 

$M_{\text{gas}}$ is set by models of how supernovae dilute metals into a mixing mass of gas. For small, early galaxies that form metal-poor stars, the mixing mass is $M_{\text{gas}} \sim 10^6 \, M_\odot$ \citep{2015MNRAS.454..659J}. The strict lower limit on this mass is the mass contained in a single final supernovae remnant, a minimum of around $\sim10^{4.5}~M_\odot$ \citep[e.g., ][]{Magg20,Macias18}, with a range of average mixing masses for metal-poor stars of $10^{5}$ to $10^{8} M_\odot$ of gas.
For systems with higher $M_{\text{gas}}$, more metals are retained, resulting in a higher retention fraction $\fret$ (and vice versa).

To estimate an average iron yield from CCSN, we calculate a weighted average between observations of H-rich CCSN and H-poor CCSN. A detailed discussion can be found in Appendix~\ref{subsec:yFe}, but find that the average yield is $y_{\text{Fe}} \approx 0.1 M_\odot$, with the uncertainty in $f_{\text{retained}}$ and $M_{\text{gas}}$ far outweighing that of $y_{\text{Fe}}$.

Altogether, $y_{\rm{Fe,eff}}$ has a wide range of possible values ($10^{-10} - 10^{-7}$), but our fiducial choice is $y_{\rm{Fe,eff}} = 10^{-9}$. This choice is validated by an independent estimation of the frequency of $r$-process events in ultra-faint dwarf galaxies in Section \ref{subsec:res_NSNfr}. We note, however, that there is tension between the values expected for $y_{\rm{Fe,eff}}$ in very low mass galaxies based on the theoretical breakdown described in this section and comparisons to several external constraints. For example, the number of supernovae predicted in an ultra-faint dwarf galaxy using the Salpeter initial mass function suggests an effective iron yield closer to $\sim10^{-7.5}$, and a simulation of extremely metal-poor ([Fe/H] $=-3.42$) stars forming after a single supernova gives an estimated effective iron yield as high as $\sim10^{-6.5}$ \citep{Chiaki19}. This is not fully unexpected as $y_{\rm{Fe,eff}}$ differs in different galaxies and the lowest mass galaxies will have the highest effective yields, but we note that this parameter remains uncertain and may trend higher than its fiducial value.

\subsection{Constraining Model Parameters: Fitting Stellar Abundance Data} \label{subsec:model_stellar}

After fixing the effective iron yield, stellar abundances are used to constrain the other model parameters. The stellar abundance data provides us effectively four observable quantities of interest:

\begin{enumerate}
\item the mean metallicity of the stars, $\avg{\text{[Fe/H]}}$, 
\item the mean $r$-process abundance, $\avg{\text{[Eu/Fe]}}$, 
\item the estimated fraction of stars that formed from gas not enriched by an $r$-process event, $\fzero$, 
\item the observed scatter in $r$-process abundance between stars, $\IQREu$. 
\end{enumerate}
Rather than quantifying scatter with standard deviation, $\sigma(\text{Eu})$, we use the more robust interquartile range, a measure of statistical dispersion equal to the difference between the 75th and 25th percentiles, denoted $\IQREu$. Model parameters are then determined as follows: 

$\avg{N_r}$ and $\alpha$: The average number of collapsars, $\avg{N_r}$, and the exponent of the $r$-process yield power law distribution, $\alpha$, are determined by comparing the observed $\fzero$ and $r$-process scatter, $\IQREu$, to those predicted by our models with varying $\avg{N_r}$ and $\alpha$. In a small metallicity bin, the shape of the [Eu/Fe] distribution function is dependent on only these two parameters. The other potentially relevant parameters contribute only to shifting the distribution to higher or lower [Eu/Fe]. By focusing on only the shape of the distribution, we can avoid making assumptions about any additional parameters when determining $\avg{N_r}$ and $\alpha$.

$N_{SN}$ and $f_r$: The number of SN enriching the gas, $N_{SN}$, is determined from the mean metallicity of the stars and the effective iron yield using $\avg{\MFe/\MH} = N_{SN} \times y_{\rm{Fe,eff}}$. The fraction of supernova that are collapsars, $f_r=\avg{N_r}/N_{SN}$, is then found by combining $N_{SN}$ and the average number of collapsars, $\avg{N_r}$, from above.

$\Mrmin$ and $\avg{M_r}$: We first determine the average $r$-process yield produced per collapsar, $\avg{M_r}$, using the relationship: $\avg{M_r} f_{\text{retained}} \XEu \approx \avg{\MEu} / \avg{N_r}$. In this equation, $\avg{\MEu}$ is found by considering the mean $r$-process abundance $\avg{\text{[Eu/Fe]}}$ and mean metallicity $\avg{\text{[Fe/H]}}$ in combination with $\MH \approx M_{\text{gas}}$. The minimum $r$-process yield is then $\Mrmin = \avg{M_r} \frac{\alpha-2}{\alpha-1}$.

We do not attempt to model higher moments of the [Eu/Fe] distribution beyond the mean and scatter because we expect selection effects in the data to dominate. We also do not attempt to model the shape of the distribution tails for both observational and theoretical reasons. Observationally, the low end of the [Eu/Fe] distribution cannot be well known without a robust selection function. Theoretically, the low and high ends of our distribution are not robust due to our assumption that model stars form after all of the supernova yields have fallen into and mixed with the hydrogen gas. This is because our assumption precludes outlier stars that, for example, could have more or less europium due to inhomogeneous mixing.

Future work will address these concerns through a more detailed treatment of enrichment that incorporates our variable-yield work into a more complete picture that includes scatter due to differences in galaxy formation (e.g., different environments or metal mixing). This will allow for a better determination of whether our assumption of a power law is an appropriate shape for the $r$-process yield distribution and improved constraints on $\alpha$ and $N_r$.

\section{Stellar Abundance Samples} \label{sec:data}

\subsection{Sample Selection}

We use a stellar abundance sample from the $R$-process Alliance (RPA), a collection of detailed abundances of 601 halo stars \citep{Hansen18,Sakari18,rpa3,RPA4}. The RPA stars are bright (V $<$ 13.5), metal-poor ([Fe/H] $\lesssim -2$) red giant stars in the Milky Way stellar halo. They were observed with a focus on obtaining a statistically complete sample of europium abundances. To verify the RPA data, we also consider a sample of 228 metal-poor red giant halo stars from \citet{Roederer14b} \citepalias[henceforth][]{Roederer14b}. Both of these samples report europium measurements or upper limits for every star.

The \citetalias{Roederer14b} sample has [Eu/Fe] abundances that are $0.22$ dex lower and [Fe/H] abundances that are $0.19$ dex lower from other samples due to using a much cooler effective temperature scale and isochrone-based surface gravities \citep{2014MNRAS.445.2946R}. We thus shift the reported measurements up by these amounts when plotting in Figure \ref{fig:starCDF} and reporting values in Table \ref{tab:CDF}.

We restrict most of our analysis to very metal-poor ([Fe/H] $< -2.5$) stars, and the highest metallicity we consider is [Fe/H] $< -1.75$ (when analyzing the evolution of the Eu scatter with increasing metallicity in Section \ref{subsec:aN}). 
We only consider stars with barium-to-europium abundance ratios that could be produced by the $r$-process ($-0.9 \lesssim $ [Ba/Eu] $ \lesssim -0.4$). [Ba/Eu] higher than $\sim -0.4$ indicates contamination from the $s$-process, another nucleosynthetic process which forms europium. The solar $r$-process barium-to-europium ratio is [Ba/Eu] $ \approx -0.8$ \citep{Sneden08}, and stars with much lower [Ba/Eu] cannot be explained by the $r$-process pattern. We note that small variations in these purity cuts do not significantly change our results.

Taking into account these restrictions (with [Fe/H] $<-2.5$), the RPA sample includes 83 stars with Eu measurements and an additional 11 stars with Eu upper limits. The \citetalias{Roederer14b} sample includes 36 stars with Eu measurements and 4 with Eu upper limits. The RPA sample (up to [Fe/H] $<-1.75$) and its $\IQREu$ in different metallicity bins can be seen in Figure \ref{fig:stellarscatter}b.

\subsection{Construction of Statistical Distributions}

\begin{figure*}[!htb]
\plotone{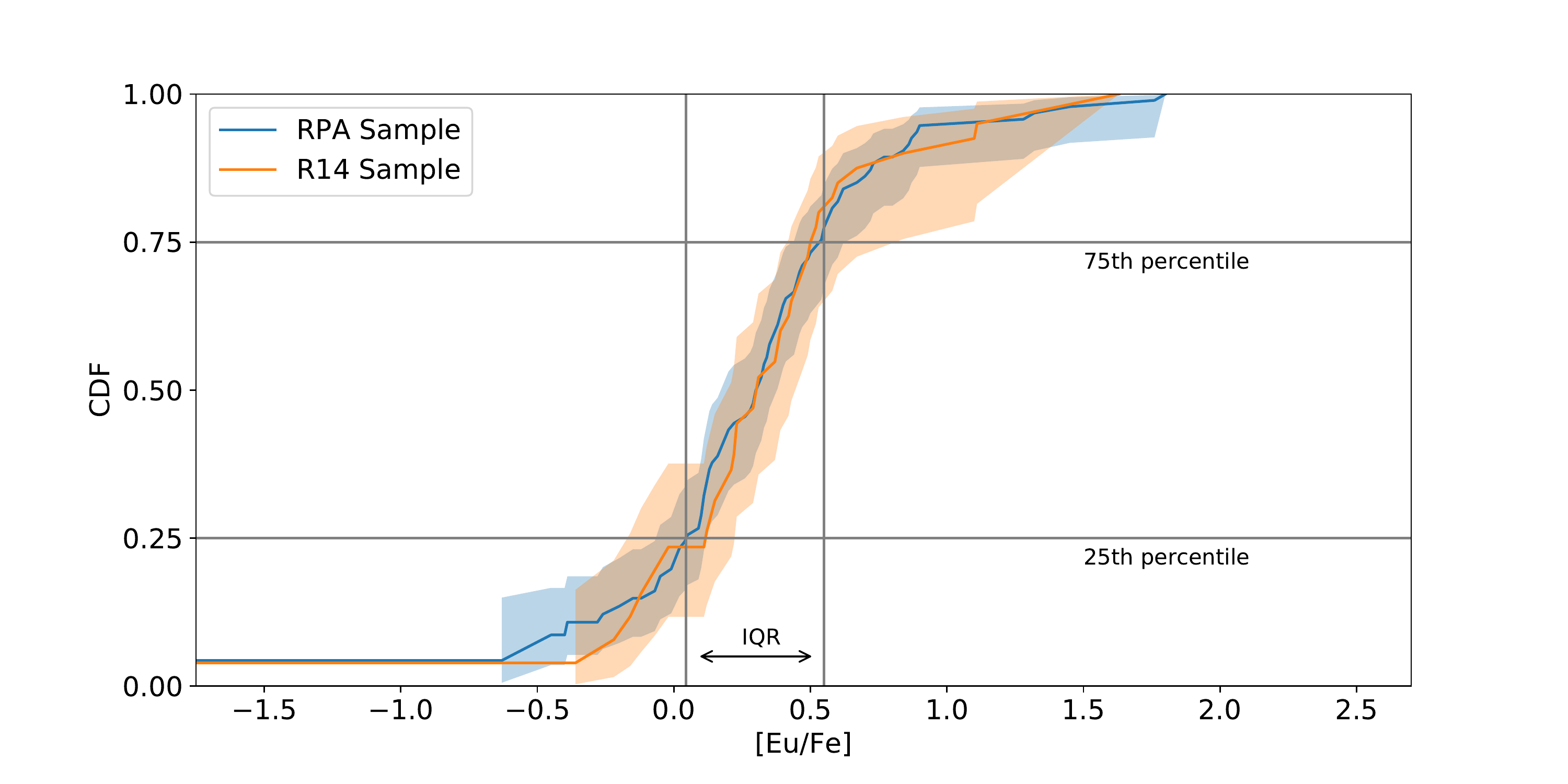}
\caption{Cumulative distribution functions for the RPA and \citetalias{Roederer14b} samples. Both CDFs are determined using the Kaplan-Meier estimator, which takes into account detections and upper limits to estimate the true distribution. The shaded regions show 95\% confidence on the CDF estimate. Grey lines outline the interquartile range (25\%-75\%) for the RPA Kaplan-Meier CDF. The CDFs have been extended to the \textit{y}-axis to show the estimated fraction of stars in each sample that have no $r$-process elements.
\label{fig:starCDF}}
\end{figure*}

\begin{table*}[!htb]
\centering
\caption{Interquartile ranges and fraction of stars formed from gas with no $r$-process enrichment for different [Eu/Fe] CDFs from observational stellar samples with [Fe/H] $< -2.5$. The distributions can be seen in Figure \ref{fig:starCDF}. $\IQREu$ uncertainties are due to both KME confidence levels and uncertain observations. The $\fzero$ values are upper limits as the distribution could continue to lower [Eu/Fe] with lower $\fzero$.
} \label{tab:CDF}
\begin{tabular}{c|c|c|c|c}
\tablewidth{0pt}
\hline
\hline
Stellar Abundance Sample & $\IQREu$ & $\fzero$ & $\avg{\text{[Eu/Fe]}}$ & $\avg{\text{[Fe/H]}}$ \\
\hline
RPA & $0.50^{+0.15}_{-0.10}$ & $0.04^{+0.10}_{-0.04}$ & $0.3^{+0.1}_{-0.1}$ & $-2.7^{+0.1}_{-0.1}$ \\
R14 & $0.38^{+0.51}_{-0.14}$ & $0.04^{+0.11}_{-0.04}$ & $0.2^{+0.2}_{-0.2}$ & $-2.9^{+0.1}_{-0.1}$ \\
\hline
\end{tabular} 
\end{table*}

To combine the mixture of measurements and upper limits into a statistical distribution of europium for each sample, we employ survival statistics, a branch of statistics that deals with censored datasets, e.g., upper limits. The most general single variate survival statistic is the Kaplan-Meier estimator (KME), which provides a non-parametric maximum likelihood estimate of a distribution from observed data. The Kaplan-Meier estimator and survival statistics have been used extensively in astronomical literature \citep[e.g.][]{upperlimits1,upperlimits2,wardle86,Simcoe04}. We use the \texttt{KaplanMeierFitter} from the survival analysis python package \texttt{lifelines} \citep{lifelines}. For this estimate to be valid, two assumptions about the distribution of upper limits must hold. First, the upper limits should be independent of each other, which is true here as the stars are independent. Second, the upper limits should be random -- i.e., the probability that a measurement will be censored should not correlate with the measurement value itself. This assumption may not hold because lower [Eu/Fe] values are more likely to be censored.
Ideally, we would fully forward model and censor our theoretical results, but that requires many additional assumptions including a completeness function (probability of measuring any value given [Eu/Fe]), an error function (the value we measure for [Eu/Fe] given its true value), and an upper limit function (the probability of setting a [Eu/Fe] upper limit at a specific value given its true value). Fully forward modeling the observational sample is beyond the scope of this paper. We thus use the Kaplan-Meier estimate while keeping in mind that this may not be a perfect estimate.

Figure \ref{fig:starCDF} shows the [Eu/Fe] cumulative distribution functions for the RPA and R14 samples. The interquartile range, $\IQREu$, differs slightly for the different samples but is consistent within the uncertainty. The mean [Eu/Fe] and [Fe/H] also differ slightly. The zero-limit $\fzero$, the estimated fraction of stars that formed from gas that was not enriched by an $r$-process event, is the same in both samples. In our model, $\fzero$ is the fraction of stars with no europium enrichment ([Eu/Fe] $= -\infty$), but we cannot identify if real stars have no $r$-process enrichment (and stars could receive trace amounts of europium enrichment through other processes despite the [Ba/Eu] cuts we applied to purify our sample). We thus estimate $\fzero$ in the data by taking the lowest CDF value from the observed distribution as estimated by survival statistics. This assumes that the CDF immediately plateaus at lower [Eu/Fe] instead of continuing to decrease. Because the distribution could continue to decrease with lower [Eu/Fe], the observed $\fzero$ values are upper limits. Realistically, the real distribution certainly does not fully plateau even if our $\fzero$ estimate is correct because of the possible other trace sources of europium, but for the purposes of this analysis and because we cannot estimate the CDF to extremely low [Eu/Fe] regardless, we ignore those minor effects. These values are shown in Table \ref{tab:CDF}.

\section{Results} \label{sec:results}

We use the stellar abundance data to constrain the model parameters. The results are summarized in Table \ref{tab:results}.

\begin{table*}[!htb]
\centering
\caption{Model parameters determined from observations. The wide ranges of $\NSN$, $\Mrmin$, and $y_{\rm{Fe,eff}}$ encompass broad uncertainty in the fraction of metals retained in each galaxy and each galaxy's gas mass. To be thorough we include these full ranges.
We also validate our fiducial values for $y_{\rm{Fe,eff}}$, $\fr$, and $\avg{N_r}$ (which also validates $\NSN$, $\Mrmin$, and $\avg{M_r}$). For $\alpha$, the full range of values produce similar distribution shapes. Derived parameter values are shown below the double line.} \label{tab:results}
\begin{tabular}{c|c|c|c}
\tablewidth{0pt}
\hline
\hline
  & Description & Range & Fiducial Value \\
\hline
$\NSN$ & 
Number of core-collapse supernovae & $30-30000$ & $3000$ \\
\hline
$\avg{N_r}$ & Average number of $r$-process collapsars & $2 - 4$ & 3 \\
\hline
$\Mrmin$ & Minimum $r$-process yield produced per collapsar (see Eq. \ref{eq:Mr}) & $3\times10^{-4} - 3\times10^{-1}$ & $3\times10^{-2}$ \\
\hline
$\alpha$ & Power law exponent of $M_r$ distribution (see Eq. \ref{eq:Mr}) & $2.2 - 6$ & $2.8$ \\
\hline
$y_{\rm{Fe,eff}}$\footnote{determined from literature values} & Effective supernovae iron yield into the total gas mass, $\yFeMgas$ & $10^{-10} - 10^{-7}$ & $10^{-9}$ \\
\hline
\hline
$f_r$ & Fraction of supernovae that are collapsars, $\avg{N_r}/\NSN$ & $10^{-4} - 10^{-1}$ & $10^{-3}$ \\
\hline
$\avg{M_r}$ & Average $r$-process yield produced per collapsar & $7\times10^{-4} - 7\times10^{-1}$ & $7\times10^{-2}$ \\
\hline
\end{tabular}
\end{table*}

\subsection{$\avg{N_r}$ and $\alpha$} \label{subsec:aN}

\begin{figure*}[!htb]
\centering
\includegraphics[width=0.875\linewidth]{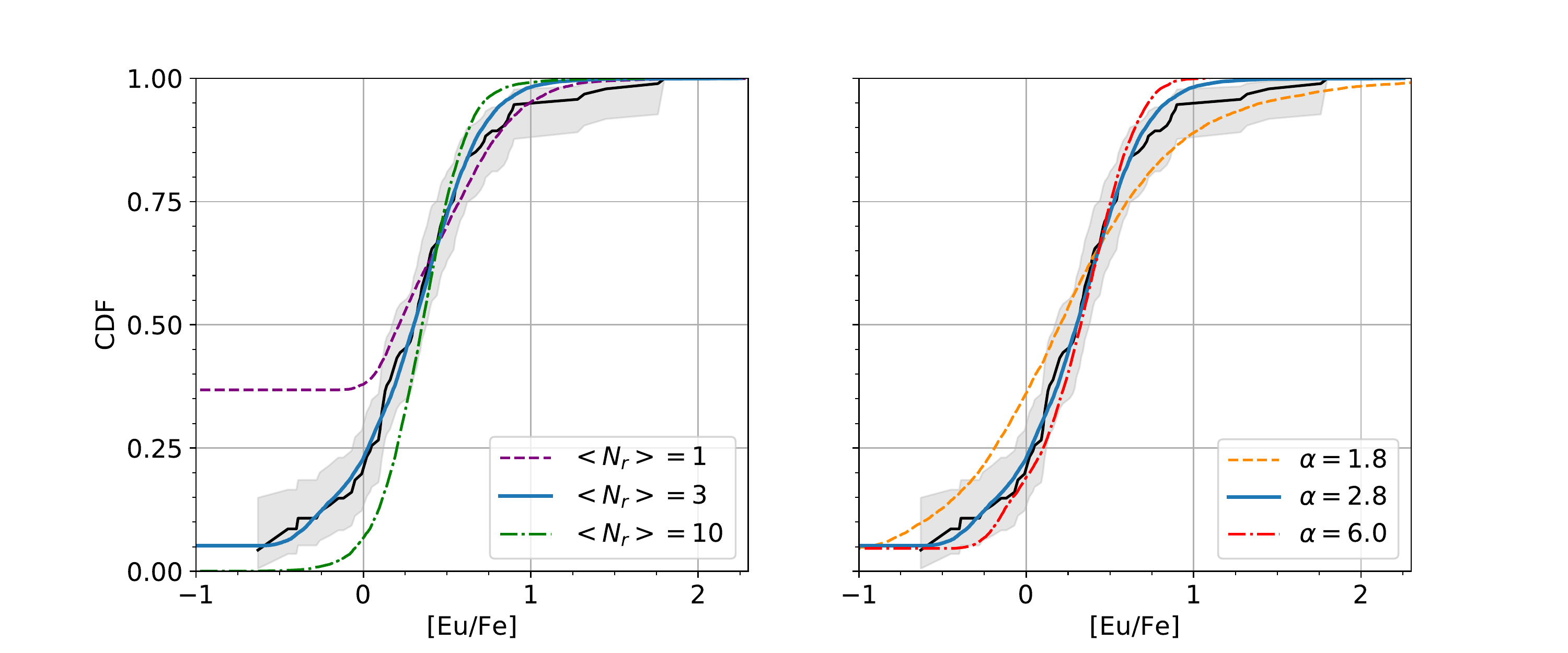}
\caption{Stellar [Eu/Fe] abundance cumulative distribution functions (colored lines) for models with different $\avg{N_r}$ and $\alpha$ values. The observed [Eu/Fe] CDF for the RPA sample is shown in black with grey uncertainty. Our fiducial model, $\avg{N_r}=3$ and $\alpha=2.8$, is shown in solid blue in both plots. When either $\avg{N_r}$ or $\alpha$ is not specified, the fiducial value is used. The model CDFs have an arbitrary offset to shift the distribution left or right for plotting purposes, so only the shape (i.e., the IQR and zero-fraction) is relevant.
\label{fig:modelCDFs}}
\end{figure*}

We use the model described in Section \ref{sec:model} to calculate theoretical cumulative distribution functions (CDF) of stellar [Eu/Fe] abundances. CDFs resulting from different representative choices of $\avg{N_r}$ and $\alpha$ can be seen in Figure \ref{fig:modelCDFs}. Each model CDF has an arbitrary offset that shifts the CDF left or right for plotting purposes. Recall that $\avg{N_r}$ and $\alpha$ can be constrained using only the shape of the distribution (i.e., the $\IQREu$ and $\fzero$). A higher $\avg{N_r}$ causes both a lower $\fzero$ since fewer stars will form from un-enriched gas and a narrower distribution due to the central limit theorem. A higher $\alpha$  also narrows the distribution by increasing the rarity of high $M_r$. When constraining these parameters with the $\IQREu$, a higher $\avg{N_r}$ thus corresponds to a lower $\alpha$ and vice versa.

\begin{figure*}[!htb]
\gridline{
	\fig{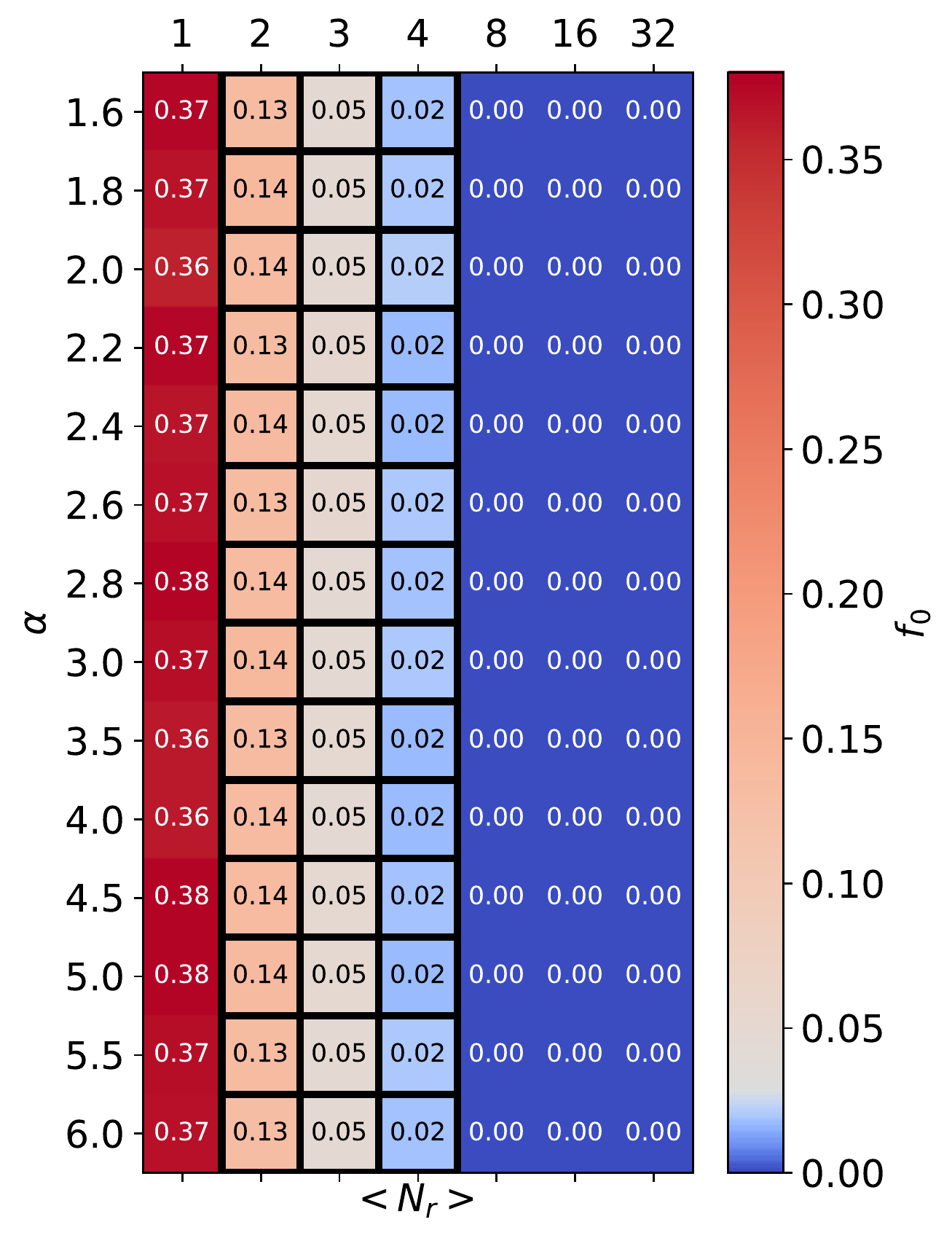}{0.4\textwidth}{(a) Values of the zero-fraction, $f_0$, for different models. The heat map colors are normalized to $\fzero = 0.04^{+0.10}_{-0.04}$, the $\fzero$ of the RPA stellar abundance sample (see Table \ref{tab:CDF}). Red is higher than observed $\fzero$, blue is lower. $\avg{N_r}$ alone affects the $f_0$ value.}
	\fig{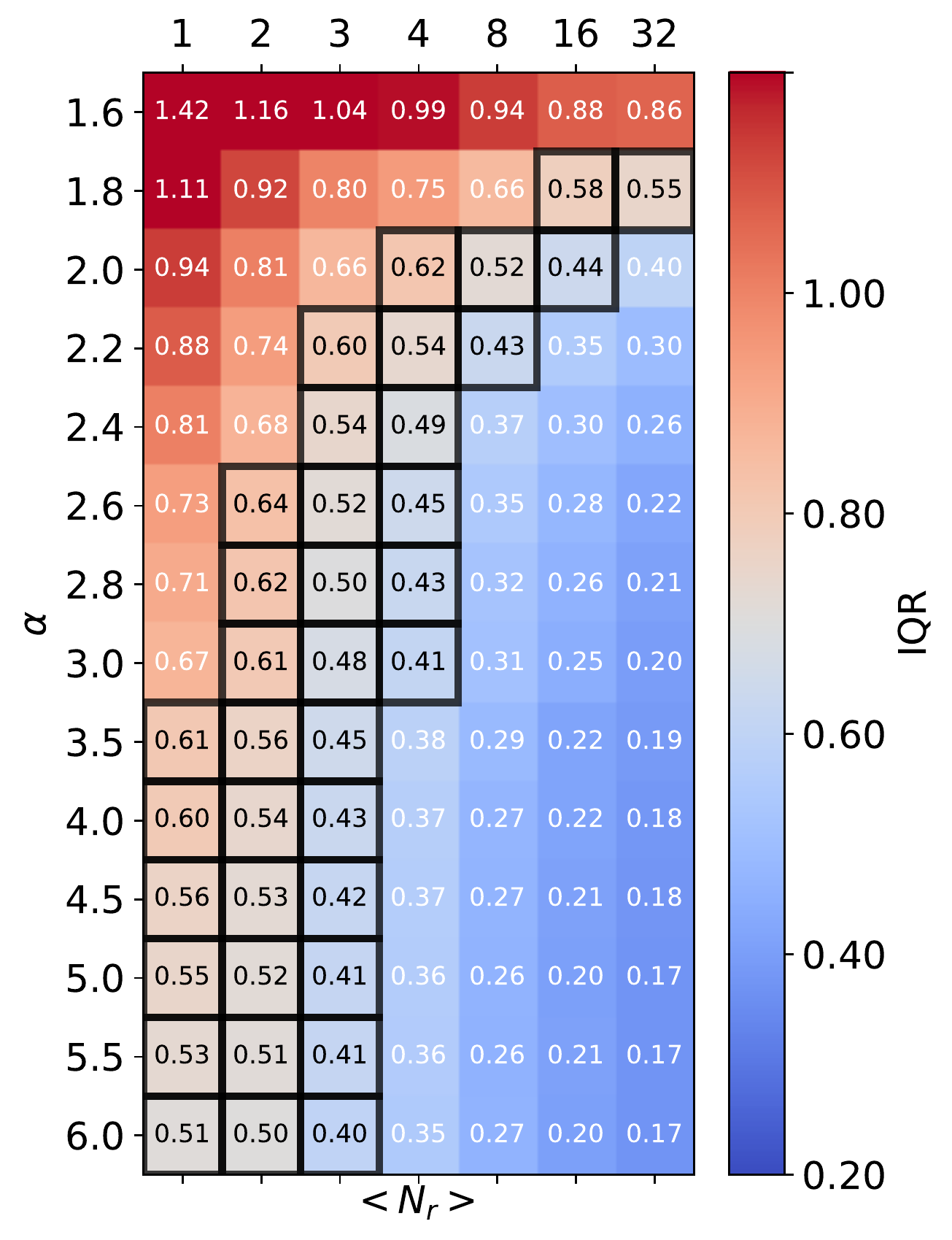}{0.4\textwidth}{(b) Values of the IQR for different models. The heat map colors are normalized to $\IQREu = 0.50^{+0.15}_{-0.10}$, the IQR of the RPA stellar abundance sample (see Table \ref{tab:CDF}). Red is higher than observed IQR, blue is lower. For $\avg{N_r}=1$, we determined the IQR by assuming symmetry and doubling the 50\%-75\% range (because the distribution is always above 25\%).}}
\caption{Heat maps showing how the cumulative distribution IQR and zero-fractions of our model vary with $\avg{N_r}$ and $\alpha$. Black boxes outline the parameter combinations that can explain the stellar data within observational uncertainty (using the RPA sample; see Table \ref{tab:CDF}). Note: The plotted $\alpha$ values increase by 0.2 until $\alpha = 3.0$, at which point they increase by 0.5 due to increasingly slower variation in $\IQREu$. \label{fig:modelHeatmaps}
}
\end{figure*}

$\avg{N_r}$, the average number of $r$-process collapsars enriching our stellar population, is constrained by the estimated fraction of stars which formed from gas not enriched by an $r$-process event, $\fzero$. For $\fzero = 0.04$, $\avg{N_r} = 3$ is the best fit value. Figure \ref{fig:modelHeatmaps}a shows how the value of $\fzero$ changes with $\avg{N_r}$, independently of $\alpha$. In this figure, the black boxes outline the parameter values which explain the observed $\fzero$ or $\IQREu$.
$\avg{N_r} = 2$ to 4 can also explain observations. Note that the observed $\fzero$ is an upper limit as the distribution could smoothly continue to lower [Eu/Fe] with a lower $\fzero$. The constraint on $\avg{N_r}$ from $\fzero$ is thus a lower bound.

\begin{figure*}[!htb]
\gridline{\fig{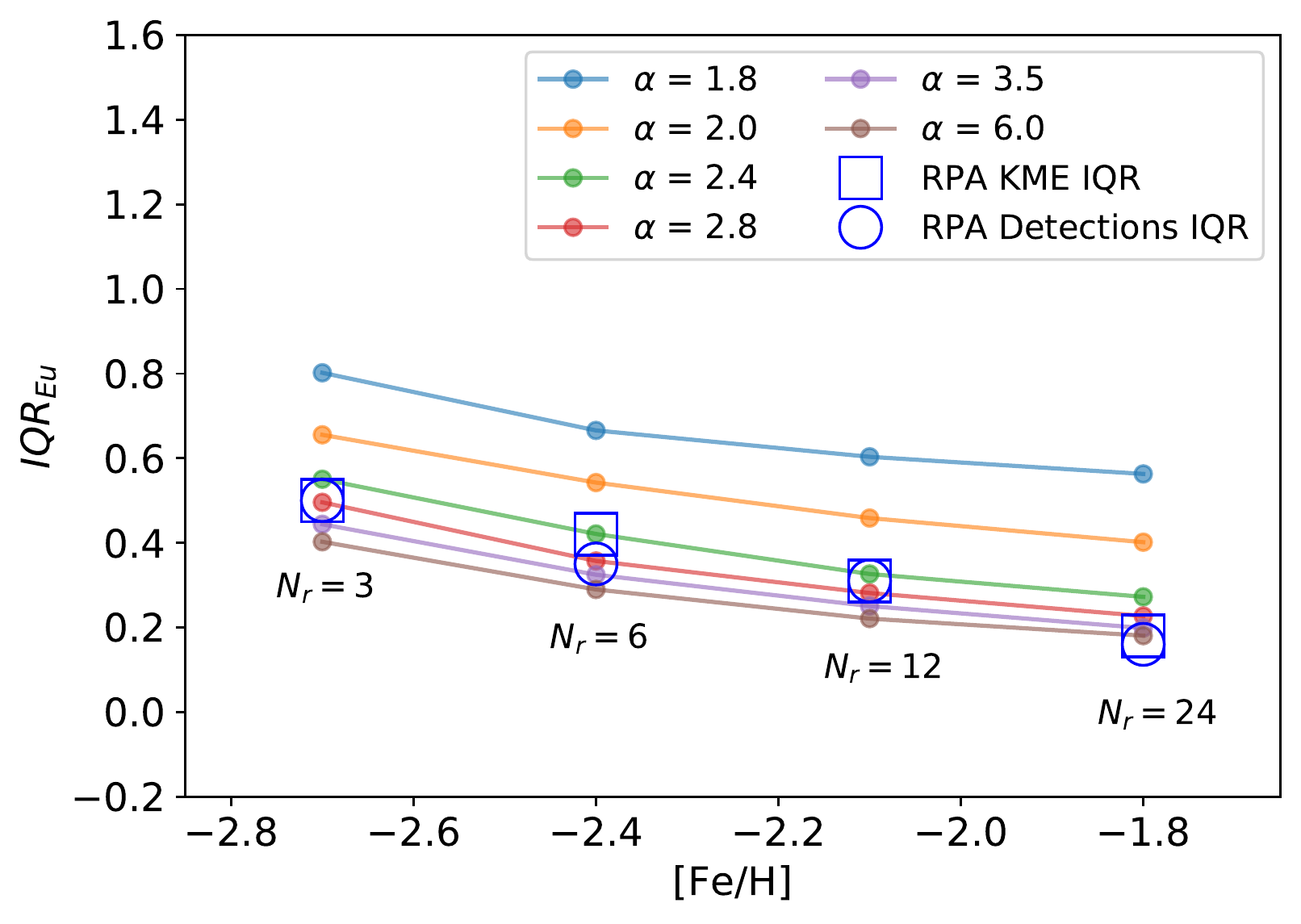}{0.45\textwidth}{(a) The evolution of the Eu scatter with increasing metallicity can be well explained by our fiducial model choices of $\avg{N_r}=3$ at $\avg{\text{[Fe/H]}} = -2.7$ (the mean metallicity of our RPA sample) and $\alpha = 2.8$.}
	\fig{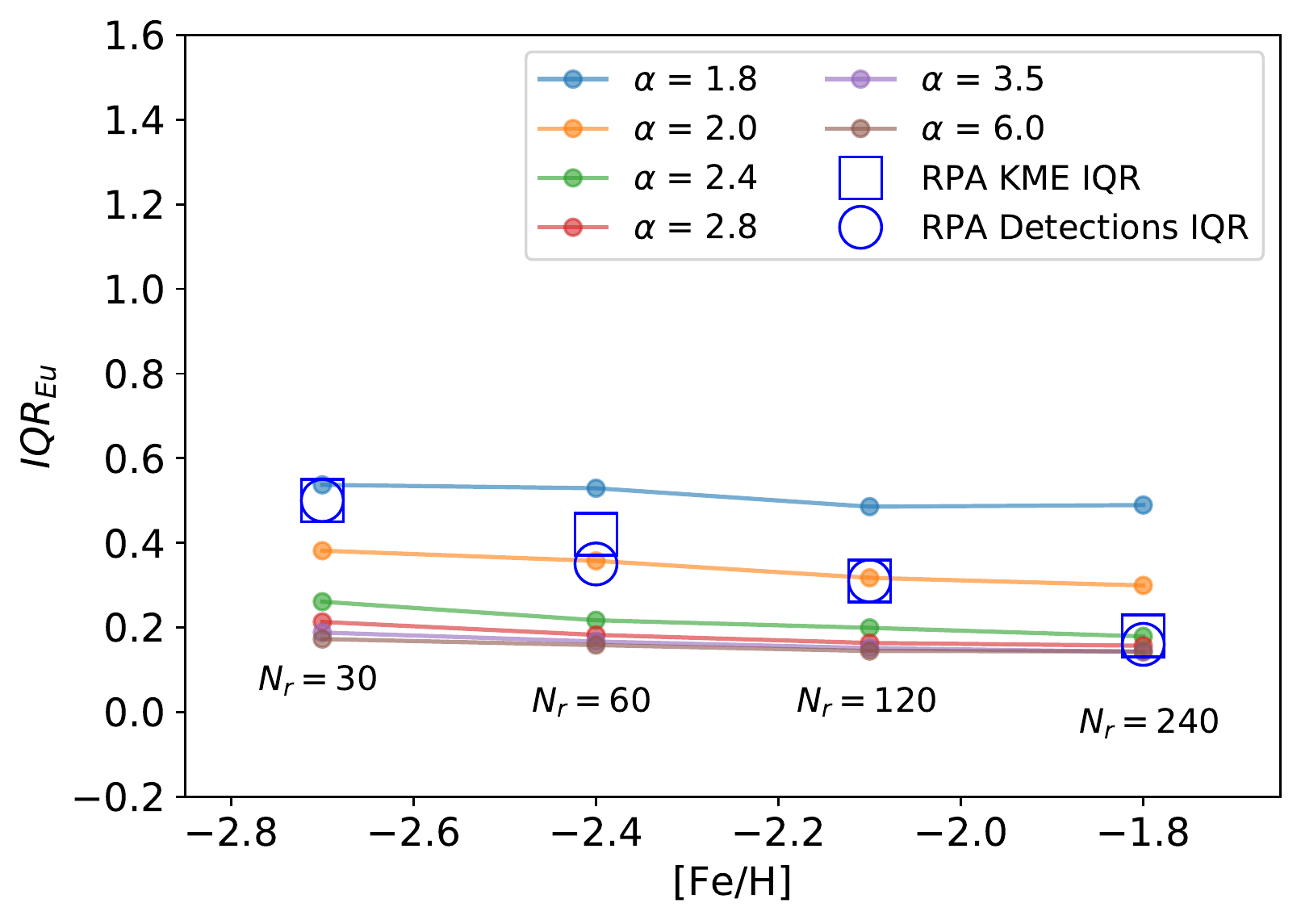}{0.45\textwidth}{(b) If $\avg{N_r}$ is increased to $\avg{N_r} = 30$ at $\avg{\text{[Fe/H]}} = -2.7$, no single choice of $\alpha$ well explains the evolution of the Eu scatter. Higher $\avg{N_r}$ choices result in poorer matches to observations.}}
\caption{The decrease in the [Eu/Fe] scatter with higher metallicity seen in the data (hollow circles and squares) is reproduced by our model (colored dots). Each [Fe/H] bin of 0.3 dex corresponds to approximately a factor of 2 increase in supernovae, hence why we double the number of $r$-process collapsars in each bin. Reproducing the evolution in the scatter at higher metallicity as well as low metallicity increases confidence in our fiducial model choices of $\avg{N_r}$ and $\alpha$. \label{fig:scatterevol}
}
\end{figure*}

To validate our fiducial value of $\avg{N_r}=3$, we also reproduce the evolution in europium scatter with increasing metallicity. This gives an upper bound to the constraint. We examine the RPA stellar abundance sample in several metallicity bins (up to [Fe/H]  $= -1.65$; see Figure \ref{fig:scatterevol}). We compare model scatter to the scatter of the RPA distribution as determined by both the Kaplan-Meier estimator (which takes into account europium detections and upper limits) and as determined by only europium detections. The \citetalias{Roederer14b} sample is excluded from this plot because it has too few stars in each bin to determine distributions.

As metallicity increases, $\avg{N_r}$ should increase linearly, but the scatter should decrease with $\sqrt{\avg{N_r}}$. Reproducing the $\IQREu$ in several metallicity bins thus suggests our model uses the correct $\avg{N_r}$. When binning on metallicity, our model with $\avg{N_r}=3$ at $\avg{[\text{Fe/H}]} =-2.7$ well reproduces the observed decrease in scatter. If we increase $\avg{N_r}$ by a factor of 10 or more, our model no longer well reproduces the observed decrease in scatter unless $\alpha$ is allowed to vary with metallicity. Considering all uncertainty from the Kaplan Meier Estimator, the upper bound is $\avg{N_r} \approx 50$, but the model favors a much lower $\avg{N_r}$. This suggests our fiducial $\avg{N_r}=3$ is roughly correct despite being a lower bound.

\begin{figure}
\includegraphics[width=0.45\textwidth]{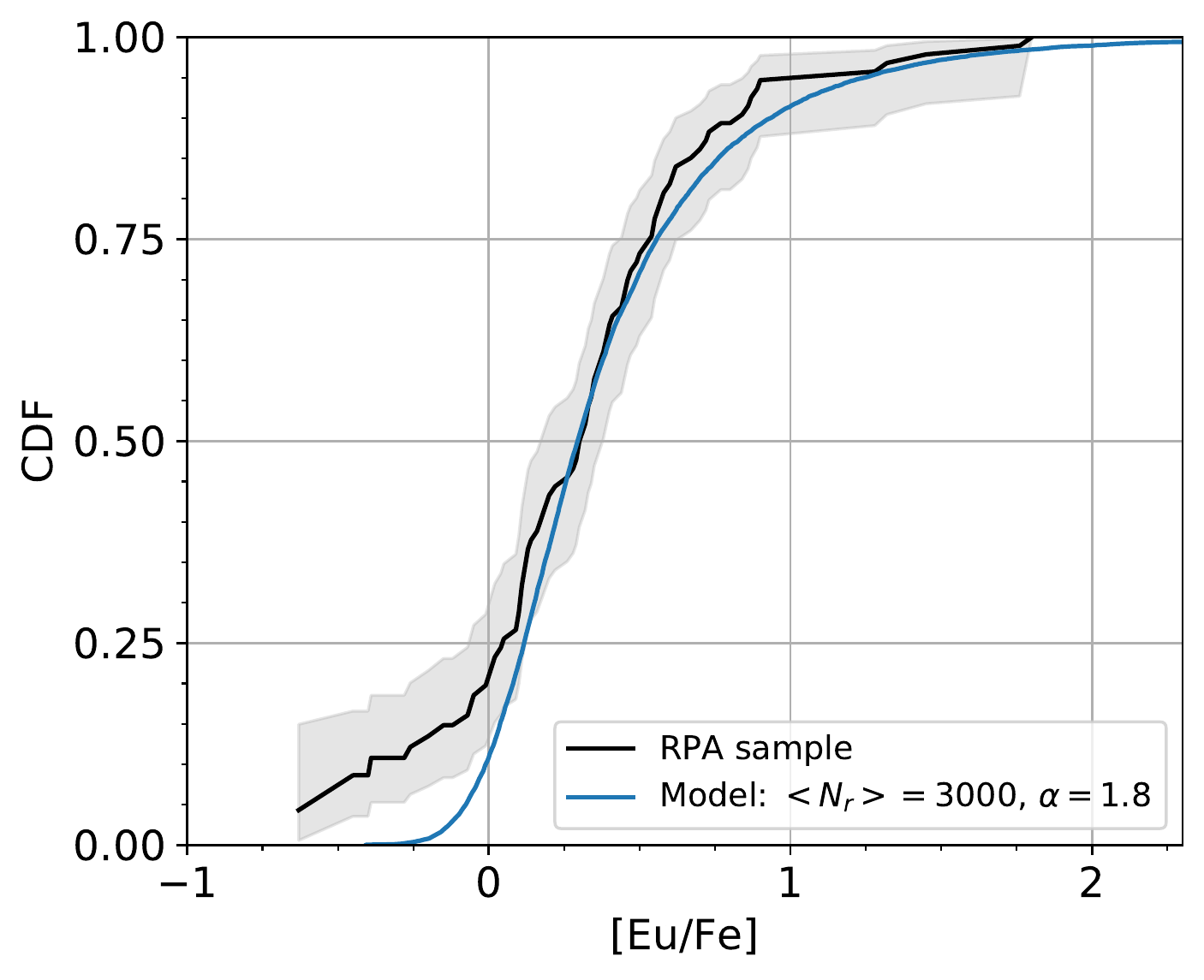}
\caption{Distribution for an extreme model where all core-collapse supernovae produce some $r$-process material ($f_r=1$). This extreme model can explain the observed $\IQREu$. It cannot well explain the evolution of scatter with increasing metallicity, however (e.g., Figure \ref{fig:scatterevol}). It also cannot well explain the low [Eu/Fe] tail. For this model, the minimum amount of $r$-process material per collapsar is extremely small, $\Mrmin \approx 10^{-6} M_\odot$. \label{fig:CDFfr1}}
\end{figure}

To be thorough, we also explore the extreme case where \textit{all} core-collapse supernovae result in $r$-process collapsars -- i.e., all core-collapse supernovae form an accretion disk that is able to synthesize a non-zero amount of $r$-process material ($\avg{N_r} = \NSN$ and $f_r=1$). We consider the case where $\avg{N_r}=3000$, where 3000 is our fiducial value of $\NSN$ (see Section \ref{subsec:res_NSNfr}). In this extreme case, the vast majority of collapsars would produce extremely small amounts of $r$-process material. As seen in Figure~\ref{fig:CDFfr1}, this extreme case can explain the observed $\IQREu$ with a scatter of $\IQREu=0.45$ for $\alpha = 1.8$. The upper bound set on $\avg{N_r}$ by the evolution of scatter with metallicity (Figure \ref{fig:scatterevol}) disfavors this model, however. The situation where all core-collapse supernovae produce $r$-process is only favored if $\NSN$ is below 30, lower than even our most extreme $\NSN$ value.
This extreme model also does not reproduce the observed distribution at low [Eu/Fe] as well as the fiducial model, though we note that the tails of the observed distribution are less trustworthy than the $\IQREu$. We thus keep $\avg{N_r}=3$ for our results.

$\alpha$, the exponent of the $r$-process yield power law distribution, is constrained by the $\IQREu$ value of the distribution, which varies with both $\alpha$ and $\avg{N_r}$ as shown in Figure \ref{fig:modelHeatmaps}b. For $\avg{N_r} = 3$ and $\IQREu = 0.50^{+0.15}_{-0.10}$, the constrained value is $\alpha = 2.8^{+4.2}_{-0.6}$.

\subsection{$\NSN$ and $\fr$} \label{subsec:res_NSNfr}

The number of supernovae, $\NSN$, is linearly related to $y_{\rm{Fe,eff}}$. To explain the mean metallicity $\avg{\text{[Fe/H]}}=-2.7 \pm 0.1$ (using the method described in Section \ref{subsec:model_stellar}), on the extreme ends of $y_{\rm{Fe,eff}}$ values we need anywhere from 30 to 30,000 supernovae; lower $y_{\rm{Fe,eff}}$ corresponds to higher $\NSN$ as more supernovae are needed to explain the mean metallicity. For our fiducial value of $y_{\rm{Fe,eff}} = 10^{-9}$, $\NSN \approx 3000$.

With values for $\avg{N_r}$ (from Section \ref{subsec:aN}) and $\NSN$, we can determine the fraction of supernovae that result in $r$-process material producing collapsars, $f_r = \avg{N_r}/\NSN$. Considering the extremes of the possible values of $y_{\rm{Fe,eff}}$, $f_r \approx 0.0001$ to $0.1$. For our fiducial values ($\avg{N_r} = 3$ and $y_{\rm{Fe,eff}} = 10^{-9}$), $f_r \approx 0.001$.

To validate our fiducial choice of $y_{\rm{Fe,eff}}$, we also estimate $f_r$ using observations of ultra-faint dwarf galaxies around the Milky Way. There are now high-resolution spectroscopic abundances for stars in 19 surviving ultra-faint dwarfs.
Of these, three of the dwarfs (Grus~II, Reticulum~II, and Tucana~III) exhibit $r$-process enrichment \citep{Hansen20_Grus,Ji16b,Hansen17}. Since these are extremely small systems, we assume each of these three dwarfs experienced one $r$-process event (as in \citealt{Ji16b,Brauer19}), and then estimate the total number of supernovae that contributed to all of their stellar populations to estimate $\fr$. We combined literature values of their absolute magnitudes $M_V$ \citep{Munoz18,Torrealba18,Drlica15,Bechtol15,Mutlu18} with a Salpeter individual mass function that predicts $0.02 L_0$ supernovae where $L_0$ is the present-day luminosity in $L_\odot$ \citep{Ji16b}. The ultra-faint dwarfs cumulatively experienced about $1800$ supernovae. The fraction of supernovae that result in $r$-process material producing collapsars is thus $\fr \sim 3/1800 = 0.002$. This validates our fiducial model values of $\fr \approx 0.001$ and $y_{\rm{Fe,eff}} \approx 10^{-9}$. We note again, however, that there is tension between our fiducial estimate of $y_{\rm{Fe,eff}}$ and several external constraints in very low mass galaxies, as discussed in Section \ref{subsec:res_yfM}, so we continue to report the full uncertainty in these parameters.

\subsection{$\Mrmin$}

The minimum $r$-process yield produced per collapsar, $\Mrmin$ (see Eq. \ref{eq:Mr}), depends on $\alpha$ and varies linearly with $y_{\rm{Fe,eff}}$. To transform between total $r$-process yield (nuclei with A $\geq 70$) and europium yield, we use the solar $r$-process europium mass fraction $\XEu \approx 10^{-3}$. To explain the observed mean europium-iron abundance ratio $\avg{\text{[Eu/Fe]}}$, on the extreme ends of $y_{\rm{Fe,eff}}$ values, we find that $\Mrmin \approx 0.0003 - 0.3 M_\odot$; lower $y_{\rm{Fe,eff}}$ corresponds to higher $\Mrmin$ both because a lower $f_{\text{retained}}$ causes less europium to be retained in the galaxy and because higher $M_{\text{gas}}$ requires a higher mass of iron and europium to explain the mean [Fe/H] and [Eu/Fe] abundances. For our fiducial values of $y_{\rm{Fe,eff}} = 10^{-9}$, $\avg{N_r} = 3$, and $\alpha = 2.8$, we find $\Mrmin \approx 0.03 M_\odot$, or a mean $r$-process yield per collapsar of $\avg{M_r} \approx 0.07 M_\odot$.

In the extreme case where all core-collapse supernovae produce a nonzero amount of $r$-process material (Figure \ref{fig:CDFfr1}), the minimum amount of $r$-process material per collapsar would be extremely small, $\Mrmin\approx 10^{-6} M_\odot$ for $\avg{N_r} \approx 3000$. This situation is disfavored because it does not reproduce the observed decrease of Eu scatter with increasing metallicity unless $y_{\rm{Fe,eff}}$ is much higher than our fiducial value.

\section{Discussion} \label{sec:disc}

Using stellar abundance data to constrain parameters in our stochastic collapsar chemical enrichment model produces a self-consistent physical picture, which was not guaranteed a priori. We now discuss this in more detail and place our results in context with other potentially physically relevant values. We also discuss the limitations of this model in Section \ref{subsec:caveats}.

\begin{figure*}[t!]
\plotone{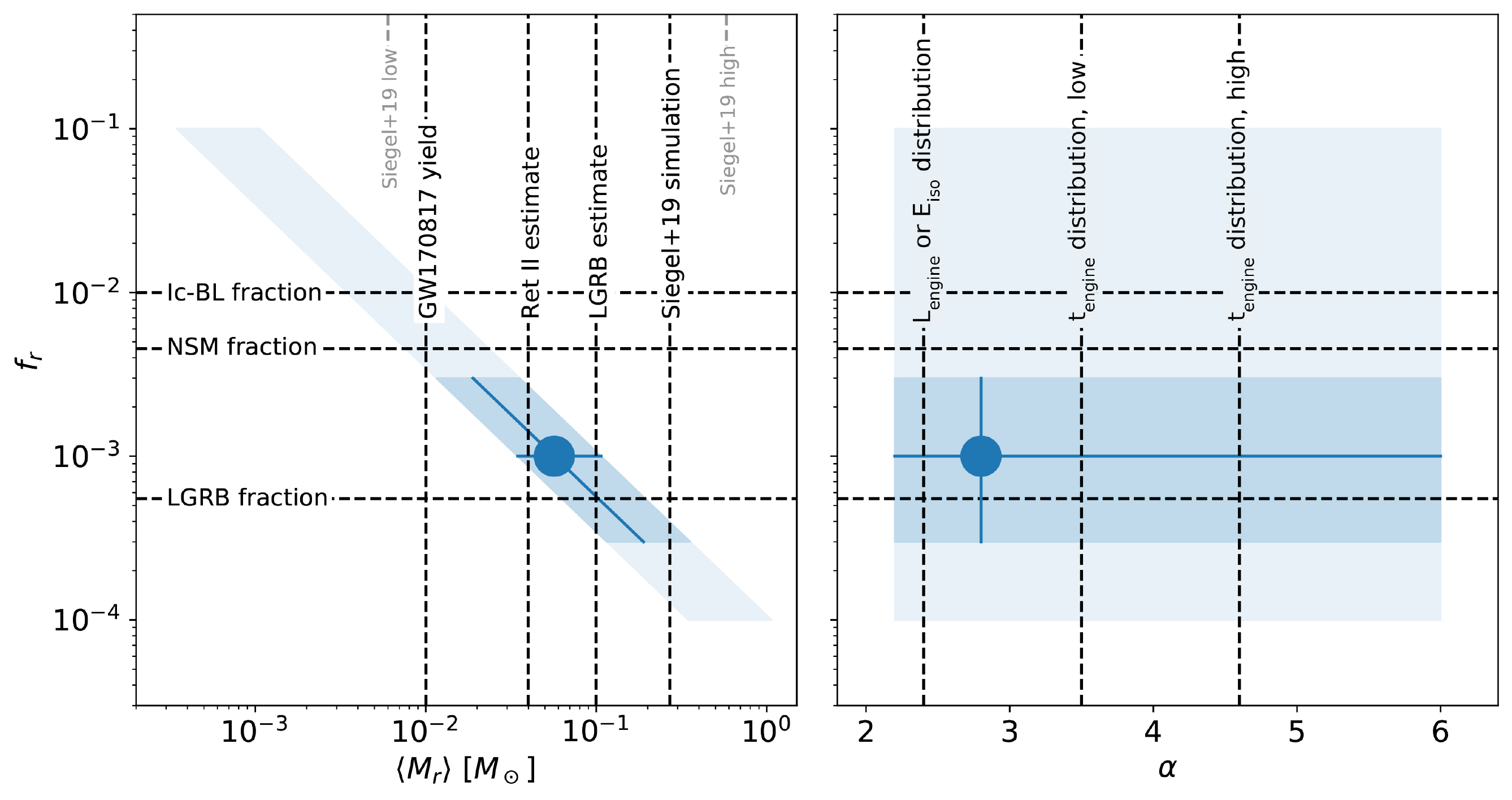}
\caption{Constraints on $\avg{M_r}$, $\fr$, and $\alpha$ from our model (blue) in context of potentially physically relevant values (black dotted lines). For descriptions of the reference values, see Section \ref{sec:disc}. Our fiducial model values are plotted as a blue dot, while the dark blue shaded region represents an order of magnitude uncertainty around our fiducial values and the light blue shaded region represents the full uncertainty.  \label{fig:constraints}}
\end{figure*}

\subsection{Implications of $f_r$: The Fraction of CCSN that Produce Collapsars} \label{subsec:disc_fr}

What fraction of core-collapse supernovae (CCSN) produce collapsars? Recall that our definition of collapsar is motivated by physical picture in which a rapid fallback accretion onto a black hole simultaneously produces heavy r-process material via accretion disk winds and launches a collimated outflow, but does \emph{not} require that a jet successfully break out of the progenitor star. The power law distribution yields adopted in Section~\ref{sec:model} naturally includes less extreme explosions that produce smaller amounts of $r$-process material via disk outflows.

To reproduce the observed scatter in [Eu/Fe] abundances at low metallicity with a collapsar-like model, we require that the fraction of CCSN producing $r$-process material is between 10$^{-4}$ and 10$^{-1}$ with a fiducial value of 10$^{-3}$. We now compare these values to the observed rates for various classes of transients that have previously been proposed to be powered by collapsars or jet-driven explosions.

We begin with long-duration $\gamma$-ray bursts (LGRBs). Current measurements of the local (z=0) rate for LGRBs beamed towards Earth from the \emph{Swift} satellite range from $1.3^{+0.6}_{-0.7}$ Gpc$^{-3}$ yr$^{-1}$ for L$>$10$^{50}$ ergs \citep{Wanderman2010} to $0.42^{+0.9}_{-0.4}$ Gpc$^{-3}$ yr$^{-1}$ accounting for complex \emph{Swift} trigger criteria \citep{Lien2014}. Correcting these for a canonical beaming factor of 50 \citep{Guetta2005}, results in inferred intrinsic rates of $65^{+0.30}_{-0.35}$ Gpc$^{-3}$ yr$^{-1}$ and $21^{+4.5}_{-2}$ Gpc$^{-3}$ yr$^{-1}$, respectively. Comparing these to the local CCSN rate from the Lick Observatory SN Search (LOSS) of 0.705 ($\pm$0.089) $\times$10$^{-4}$ Mpc$^{-3}$ yr$^{-1}$ \citep{Li2011} yields R$_{\rm{LGRB}}$/R$_{\rm{CCSN}}$ values of (2$-$9)$\times$10$^{-4}$. This range of values is only slightly lower than our fiducial value of $f_r$. LGRBs could thus be linked to $r$-process production. Our uncertainty on $f_r$ errs toward a higher value for the $r$-process fraction, though, so $f_r$ could very well be larger than $f_{\text{LGRB}}$. In that case, we would require that massive stars beyond those that launch successful GRBs form accretion disks with physical conditions capable of producing heavy $r$-process material.

In particular, Type Ic-BL supernovae are a class of hydrogen-poor SN that display high ejecta velocities (hence ``broad-lined'') and kinetic energies ($\sim$10$^{52}$ ergs) for which central engines are commonly evoked. While the nature of the central engine is still debated \citep{Thompson2004,MacFadyen99,Barnes2018}, the fact that all SN observed in association with LGRBs have been Type Ic-BL supernovae has lead to the hypothesis that all events of this class are powered by jets. Differences in the detailed manifestation of these explosions (LGRBs, low-luminosity GRBs, relativistic SN, or ``ordinary'' Ic-BL SN) would then be driven by a distribution of engine timescales or progenitor radii \citep[e.g.][]{Lazzati2012,Margutti2014}. We therefore compare our constraints on $f_r$ to the rates of Type Ic-BL SN to investigate if they are consistent with all Type Ic-BL SN harboring collapsar engines.

Based on the full LOSS sample, \citet{2017PASP..129e4201S} find that Type Ic-BL SN account for a fraction of (1.1 $\pm$ 0.8) $\times 10^{-2}$ of CCSN.  However, the LOSS sample was a targeted survey, biased towards high metallicity galaxies and it is well established that Type Ic-BL SN show a preference for low metallicity environments \citep[e.g.][]{Modjaz2020}. It is therefore possible that Type Ic-BL SN represent a higher fraction of all CCSN at low metallicity, which is what our parameter $f_r$ actually constrains. Unfortunately, to date, there has been no untargeted, volume-limited study that examines the fraction of CCSN that are Type Ic-BL at low-metallicity. \citet{Graur2017b} and \citet{Arcavi2010} examine relative rates of different core collapse SN subtype in ``high'' and ``low'' mass galaxies for the LOSS and early PTF samples, respectively. \citet{Graur2017b} find no significant difference in the Type Ic-BL fraction (1-2\%), while \citet{Arcavi2010} find that Type Ic-BL may make up a significantly higher fraction of all SN ($\sim 10-13\%$) in low luminosity galaxies. We caution, however, that both samples contain only 2-3 Type Ic-BL events and are therefore dominated by low number statistics. More recently, \citet{2020arXiv200805988S} investigate the host galaxies of the full sample of 888 SN identified by PTF, including 36 Type Ic-BL. They find that Type Ic-BL production is significantly stifled above a galaxy mass of $\log{M/M_\odot} = 10$, with Type Ic-BL comprising $\gtrsim 5$\% of their observed CCSN sample below this threshold compared to $\lesssim 2$\% above.

R$_{\rm{Ic-BL}}$/R$_{\rm{CCSN}}$ values of 0.01--0.1 fall within the range of $f_r$ found by our model (see Figure~\ref{fig:constraints}). However, the latter is at the extreme high end, implying that while our model is consistent with all Type Ic-BL SNe producing europium, it favors a scenario in which $\lesssim$10\% do. We note that this would not preclude the possibility that all Type Ic-BL SNe harbor jets, but rather require that some lack the accretion disk properties necessary for the production of \emph{heavy} r-process material. This could imply that a subset of Type Ic-BL SNe (a) harbor accreting black holes, but do not reach sufficiently high accretion rates ($> 10^{-3}$ M$_\odot$ s$^{-1}$) to proceed past $^{56}$Ni-rich outflows \citep{siegel19}, or (b) harbor magnetar central engines for which neutrino irradiation can limit neucleosynthesis from disk ejecta to the light r-process \citep[e.g.][]{Margalit17,Radice18}.

For comparison, we also calculate a rough effective rate of neutron star mergers per CCSN. The cosmic NSM rate from the second LIGO-Virgo gravitational wave transient catalog is $320^{+490}_{-240}$ Gpc$^{-3}$ yr$^{-1}$ \citep{2020arXiv201014533T}. When comparing this to the LOSS rate of galactic CCSN, the estimated NSM fraction is $4.5^{+7.0}_{-3.4} \times 10^{-3}$. This is higher than our fiducial value of $f_r$, but within model uncertainties. The LIGO rate of NSMs could thus potentially account for the rate of $r$-process events required by our model to explain metal-poor star abundances, though it is not favored by our fiducial results. We also note that the rate of NSMs in the early universe likely differs from the rate found by LIGO, and that NSMs would need to be fast-merging to be described by our model.


\subsection{Implications of $\avg{M_r}$: The Amount of $r$-Process Yield Produced per Collapsar}

Our determination of the minimum and average amounts of $r$-process yield produced per collapsar ($\Mrmin \approx 0.03 M_\odot$ and $\avg{M_r} \approx 0.07 M_\odot$, respectively) is based entirely on our analysis of RPA stellar abundance data, independent of any previous estimates in literature of the amount of $r$-process material that might be produced by such events. To place our results in context, we compare them to several reference estimates of $r$-process yields from single events (see Figure \ref{fig:constraints}). Note again that we define $r$-process yield as the yield of nuclei with mass number A $\geq 70$.

\citet{siegel19} demonstrated that accretion disk outflows in collapsars could produce significant amounts of $r$-process material. For different presupernova models, they found the amount of europium varied from $6.0 \times 10^{-6} \, M_\odot$ to $5.8 \times 10^{-4} \, M_\odot$, or $M_r = 0.006$ to $0.579 \, M_\odot$ (for our definition $M_r$). Their fiducial model corresponds to $M_r = 0.27  M_\odot$. Their fiducial yield is about four times larger than our fiducial average yield, but our $\Mrmin$ and $\avg{M_r}$ values fall within their range of yields.
In Figure \ref{fig:constraints}, the shaded region correspond to the spread of $r$-process yields found by the \citet{siegel19} simulations.

Furthermore, if we assume that the isotropic energy of a $\gamma$-ray burst roughly traces the amount of $r$-process yield, we can compare the energies of LGRBs to that of GW170817 to estimate the $M_r$ from collapsars in which the associated jet successfully breaks out of the progenitor star. This assumption predicates on the ideas that (1) the same physical processes act in both short and long GRBs and (2) the accretion phase during which europium is produced roughly coincides with the phase during which the GRB occurs in the source frame, matching assumptions of \citet{siegel19}. \citet{cote18} infer that $\sim 3-15 \times 10^{-6} M_\odot$ of europium was ejected from the post-merger accretion disk of GW170817. This translates to $\sim 0.01 M_\odot$ of heavy $r$-process material for a europium mass fraction of $\XEu$~$=$~$10^{-3}$.
The istropic $\gamma$-ray energy of GW170817 was $E_{\gamma,iso,GW170817} = 2.1^{+6.4}_{-1.5} \times 10^{52}$ ergs \citep{GW170817Eiso}, and from a sample of 468 LGRBs, the mean istropic energy of LGRBs is $E_{\gamma,iso,LGRBs} \approx 2.6^{+2.7}_{-0.5} \times 10^{53}$ ergs \citep{LGRB_stats}. With these values:
$$M_{r,collapsar}\sim M_{r,GW170817}\frac{E_{\gamma,iso,LGRBs}}{E_{\gamma,iso,GW170817}} \sim 0.1 M_{\odot}$$ 
(see also \citealt{Siegel20}). This aligns with our fiducial value of $\avg{M_r}$. In particular, our fiducial results lie near the intersection of the $r$-process yield expected per LGRB and the fraction of LGRBs per CCSN (see Figure \ref{fig:constraints}). This supports the possibility that LGRBs are linked to $r$-process production.

For the final reference mass, we compare to the amount of $r$-process yield that was produced in the $r$-process event that enriched the ultra-faint dwarf galaxy Reticulum II \citep{Ji16b}. This galaxy preserves $r$-process enrichment from a single prolific event in the early universe. To explain the europium abundances of its stars, it likely experienced an event with a europium yield of $10^{-4.3}$ to $10^{-4.6} M_\odot$ \citep{Ji16b}. With $\XEu=10^{-3}$, this corresponds to $M_r \sim 0.04 M_\odot$. Our $\avg{M_r}$ value is only slightly higher than this mass. This yield is also consistent with that expected for neutron star mergers \cite[e.g., the yield estimated from GW170817,][]{siegel2019b,cote18}.

\subsection{Implications of $\alpha$: Learning About Collapsar Properties from $r$-Process Abundance Scatter} \label{subsec:disc_a}

Unfortunately, the current precision on the shape of the [Eu/Fe] distribution does not provide tight constraints on $\alpha$, the exponent of our $r$-process yield power law distribution. For $\avg{N_r} = 3$, any $\alpha = 2.2-6.0$ can  explain the observed scatter. Our fiducial value of $\alpha = 2.8$ best fits the data, but the full range of possible values produces similar distribution widths (see Figure \ref{fig:modelHeatmaps}b).

The $\alpha$ constraints from metal-poor stars can be compared to power law distributions of long $\gamma$-ray burst (LGRB) engine duration, engine luminosity, and isotropic energy. Figure \ref{fig:constraints} shows our constraints on $\alpha$ in context with the exponents from these distributions.

\citet{petropoulou17} modeled the central engines which power LGRBs, determining power law distributions for both the engine luminosities and engine activity times:  $p(L_{engine})\propto L^{-\alpha_L}$ and $p(t_{engine})\propto t^{-\alpha_t}$. By assuming that more powerful engines can more quickly break out of the collapsing star to produce $\gamma$-ray signals (with a breakout time that scales with jet luminosity as $L^{-\chi}$), they show that the shape of the $\gamma$-ray duration distribution can be uniquely determined by the observed GRB luminosity function. In particular, they determine the power law indexes of the $L_{engine}$ and $t_{engine}$ distributions by connecting them with the observed distributions of luminosities and durations of LGRBs. For $\chi=1/3$, \citet{petropoulou17} find $\alpha_L=2.4$ and $\alpha_t=3.5$, while for $\chi=1/2$, they constrain $\alpha_L=2.4$ and $\alpha_t=4.6$. In addition, by assuming a single breakout time, \citet{sobacchi17} find a power law distribution for $t_{engine}$ consistent with $\alpha_t\sim4$. 

Furthermore, we can determine the isotropic energy distribution of LGRBs since $E \propto L \times t$. Because both $L_{engine}$ and $t_{engine}$ draw from power law distributions, the distribution of their product follows the distribution of the variable with a smaller $\alpha$, in this case $\alpha_L = 2.4$.

\begin{figure}
\includegraphics[width=0.45\textwidth]{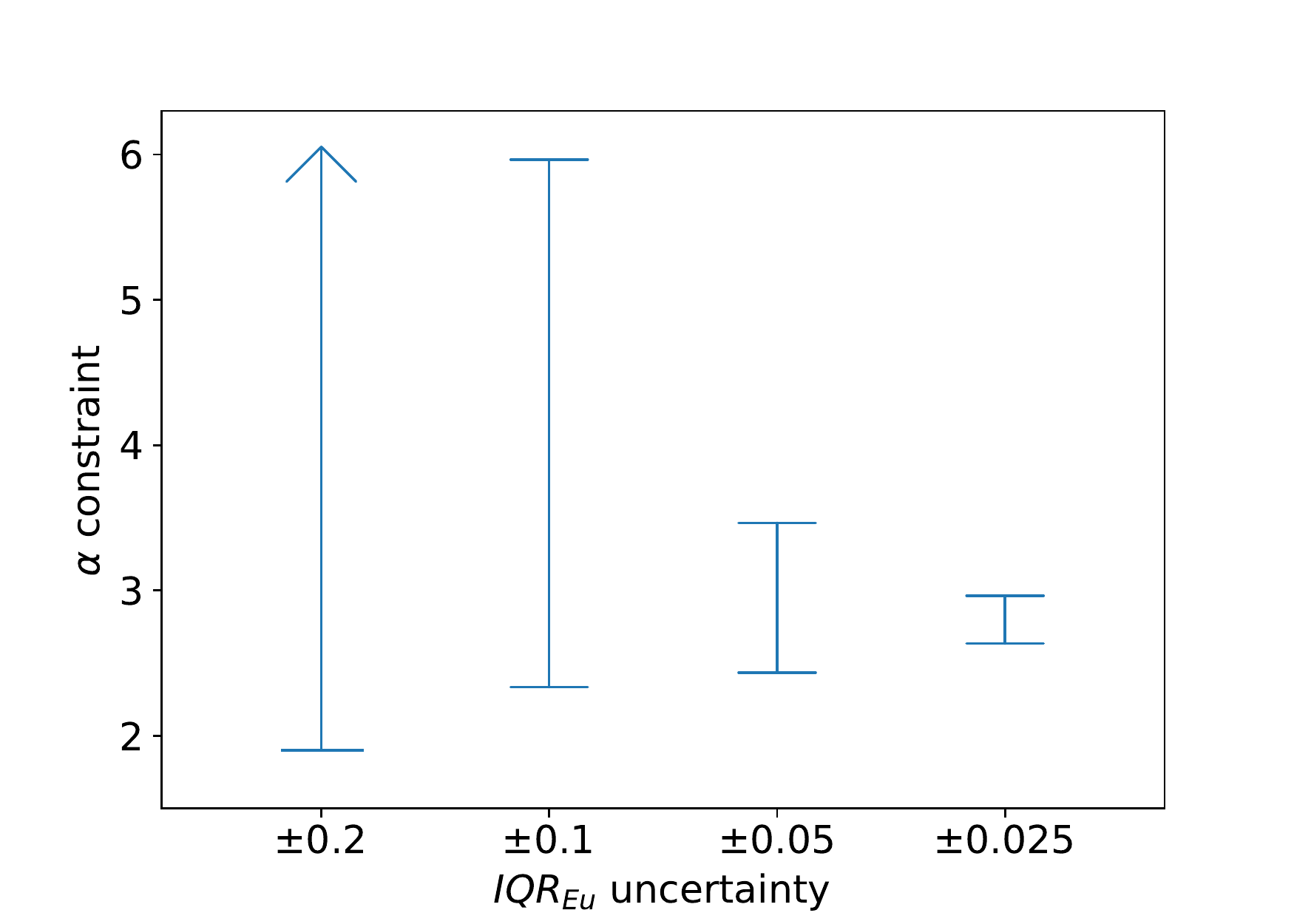}
\caption{To improve our constraint on $\alpha$, we must improve our measurement of the stellar $\IQREu$ for metal-poor stars. Here we show how the $\alpha$ constraint is improved for several different $\IQREu$ uncertainties. \label{fig:IQRuncertainty}}
\end{figure}

Our $\alpha$ constraint overlaps with all of these values, with the fiducial value falling closer to $L_{engine}$ or $E_{iso}$. Any of these properties could therefore potentially trace the $r$-process yield.
For a better constraint on $\alpha$, we need a significantly lower uncertainty on the observed $\IQREu$. Figure \ref{fig:IQRuncertainty} shows how tightly $\IQREu$ must be measured for the stellar samples to improve the $\alpha$ constraint. This plot was constructed assuming the $\IQREu$ is centered on $\IQREu=0.50$, as found for the RPA sample. To differentiate between the distributions for $t_{engine}$ and $L_{engine}$ or $E_{iso}$, the $\IQREu$ must be measured with uncertainty $<0.05$ dex. This abundance precision is better than what current measurements can achieve in metal-poor stars, though it may become achievable in the future as stellar spectroscopy methods improve.

The $\IQREu$ is a robust but very inefficient estimator of the distribution shape. Alternatively, we could use the full distribution shape. This requires a reliable selection function, but would likely not demand 0.05 dex precision.

\subsection{Neutron Star Mergers vs. Collapsars} \label{subsec:NSMvscollapsar}

Here we focused on collapsars, demonstrating that a $r$-process production site with a power law distribution inspired by LGRB jet properties can self-consistently reproduce the abundances and scatter observed in metal-poor stars. These results would also apply to another prompt $r$-process site ejecting a Solar $r$-process abundance pattern that scales with a power law, however.

The possibility that collapsars produce $r$-process material is debated. For example, earlier semi-analytic work on collapsar disk winds by \citet{Surman06} found that collapsar outflows are too neutron-poor to produce heavy $r$-process isotopes. A recent study by \citet{Miller19} that investigated the \citet{siegel19} results with more detailed modeling of neutrino transport also found that collapsar outflows are incapable of producing third peak $r$-process material. Furthermore, \citet{Macias19} found that any $r$-process site that also produces large amounts of iron is disfavored by observations of metal-poor stars. Collapsars that do not produce large amounts of iron (e.g., LGRBs without an associated supernovae; \citealt{2006Natur.444.1047F}) would avoid the dilution problems discussed by Macias \& Ramirez-Ruiz, but the topic is unsettled.

Neutron star mergers are a demonstrated source of $r$-process thanks to GW170817 and, in principle, their europium yields can vary as well with different neutron star binary masses, mass ratios \citep{Korobkin12,Bauswein13,Hotokezaka13,dietrich15,Sekiguchi16}, and eccentricities \citep{Chaurasia18,Papenfort18}.
Furthermore, the $\Mrmin$ and $\avg{M_r}$ values in our model are roughly consistent with the $r$-process yield estimated for GW170817 \citep{siegel2019b,cote18}. Because of this, variable-yield neutron star mergers could also potentially explain the $r$-process scatter in metal-poor stars via a similar model to that presented in this paper. More work is needed to determine a reasonable distribution of $r$-process effective yields from neutron star mergers, combining input distributions of binary neutron star properties and yields (e.g., those from the numerical simulations cited above) and kick velocities \cite[e.g.,][]{Tarumi20, Safarzadeh19b, Bonetti19}.

\subsection{Limitations of Initial Model} 
\label{subsec:caveats}

Our initial model is purposefully simple in order to act as a focused exploration of variable-yield collapsars. In particular, the model assumes all abundance scatter is due to variable stellar populations. This assumption allows us to expressly investigate variable yields as a source of scatter, but it does not consider possible effects due to differences in galaxy formation. Real dwarf galaxies have differences in their hierarchical assembly, small amounts of cross-pollution, and experience inhomogeneous mixing \cite[e.g.,][]{Venn04,Ji15,Griffen18}. Abundance scatter is likely affected by these complexities. Inhomogeneous enrichment has been included in some previous models \cite[e.g.,][]{Cescutti15,Wehmeyer15}, but it is not a solved problem. In this very metal-poor regime, theoretical work has not yet given a simple way to model the amount of scatter from galaxy formation effects.

This model also assumes that each star probes an independent gas reservoir. For every star in our model, we assume that it originates from a different dwarf galaxy in which a number of SN exploded over some time, the metals fell back down into the galaxy and fully mixed, and then our model star formed from the mixed gas. This approximates the average star that formed in a given gas reservoir. Real stellar samples likely contain stars that originated together, though. Observational work that studies the accretion origin of stars through, for example, analysis of stellar streams and kinematic clustering will inform the quality of this assumption in the future.

To transcend the limitations of this initial model, future models will consider scatter due to differences in galaxy formation and include a more detailed treatment of chemical enrichment and star formation. We are currently developing high-resolution hydrodynamic simulations of dwarf galaxy evolution that will study these effects and further explore the origins of $r$-process material.

\section{Conclusions} \label{sec:conc}

We have produced a self-consistent model in which collapsars synthesize all of the $r$-process material in the early universe. By assuming the $r$-process material in metal-poor ([Fe/H] $< -2.5$) stars was formed exclusively in collapsars with stochastic yields, we can reproduce the observed distribution of europium abundances with parameter values that are consistent with other independently determined reference values. This was not guaranteed a priori.

This is not evidence that collapsars dominantly produce $r$-process material in the early universe, however. Neutron star mergers with variable effective europium yields may also be able to explain the $r$-process scatter. More work is needed on the effective europium yields of neutron star mergers. In particular, the retention fraction is important for the collapsar model, but it becomes even more important for neutron star mergers with different natal kick velocities and coalescence times.

Abundance scatter of metal-poor stars is an important window into the different mechanisms producing $r$-process elements. Individual mechanisms can produce scatter without the need for multiple sources. In this paper, we assume a power law distribution of collapsar $r$-process yields. The range of constrained values for the exponent $\alpha$ is comparable to those of the distributions for long $\gamma$-ray burst isotropic energies, engine luminosities, and engine times (Section \ref{subsec:disc_a}). Improved constraints on $\alpha$ could allow us to investigate which, if any, of these collapsar properties trace $r$-process yield. 

Lastly, in our model, the fraction of core-collapse supernovae that result in $r$-process collapsars, $f_r$, is comparable to the fraction of core-collapse supernovae that result in long $\gamma$-ray bursts. This could indicate a link between LGRBs and $r$-process. The uncertainty in our model errs to higher $f_r$, though, and if $f_r$ is higher then we would require a significant number of $r$-process collapsars that do not produce long $\gamma$-ray bursts. Our model also favors a scenario in which $\lesssim 10$\% of Type Ic-BL supernovae produce europium. This does not preclude all Ic-BL SNe from harboring choked jets, but would imply that some Ic-BL SNe lack the accretion disk properties to synthesize heavy $r$-process isotopes.

\acknowledgments
We thank Daniel Siegel for providing nucleosynthesis products from his group's collapsar simulations and for helpful discussions.
We also thank Paz Beniamini Tony Piro, 
Enrico Ramirez-Ruiz, Phillip Macias, and Anne Kolborg for helpful discussions.
KB acknowledges support from the United States Department of Energy grant DE-SC0019323.
APJ acknowledges support by NASA through Hubble Fellowship grant HST-HF2-51393.001, awarded by the Space Telescope Science Institute, which is operated by the Association of Universities for Research in Astronomy, Inc., for NASA, under contract NAS5-26555.
APJ also acknowledges a Carnegie Fellowship and the Thacher Research Award in Astronomy.
MRD acknowledges support from the NSERC through grant RGPIN-2019-06186, the Canada Research Chairs Program, the Canadian Institute for Advanced Research (CIFAR), and the Dunlap Institute at the University of Toronto.
AF acknowledges support from NSF grant AST-1716251, the Silverman (1968) Family Career Development Professorship and thanks the Wissenschaftskolleg zu Berlin for their wonderful Fellow's program and generous hospitality.

\bibliography{kaleybib,alexbib}{}

\begin{thebibliography}{}
\expandafter\ifx\csname natexlab\endcsname\relax\def\natexlab#1{#1}\fi
\providecommand{\url}[1]{\href{#1}{#1}}
\providecommand{\dodoi}[1]{doi:~\href{http://doi.org/#1}{\nolinkurl{#1}}}
\providecommand{\doeprint}[1]{\href{http://ascl.net/#1}{\nolinkurl{http://ascl.net/#1}}}
\providecommand{\doarXiv}[1]{\href{https://arxiv.org/abs/#1}{\nolinkurl{https://arxiv.org/abs/#1}}}

\bibitem[{{Abbott} {et~al.}(2017){Abbott}, {Abbott}, {Abbott}, {Acernese},
  {Ackley}, {Adams}, {Adams}, {Addesso}, {Adhikari}, {Adya}, \&
  et~al.}]{LIGOGW170817a}
{Abbott}, B.~P., {Abbott}, R., {Abbott}, T.~D., {et~al.} 2017, Physical Review
  Letters, 119, 161101, \dodoi{10.1103/PhysRevLett.119.161101}

\bibitem[{{Afsariardchi} {et~al.}(2020){Afsariardchi}, {Drout}, {Khatami},
  {Matzner}, {Moon}, \& {Ni}}]{Afsariardchi2020}
{Afsariardchi}, N., {Drout}, M.~R., {Khatami}, D., {et~al.} 2020, arXiv
  e-prints, arXiv:2009.06683.
\newblock \doarXiv{2009.06683}

\bibitem[{{Anderson}(2019)}]{Anderson19}
{Anderson}, J.~P. 2019, \aap, 628, A7, \dodoi{10.1051/0004-6361/201935027}

\bibitem[{{Andrews} {et~al.}(2020){Andrews}, {Breivik}, {Pankow}, {D'Orazio},
  \& {Safarzadeh}}]{Andrews20}
{Andrews}, J.~J., {Breivik}, K., {Pankow}, C., {D'Orazio}, D.~J., \&
  {Safarzadeh}, M. 2020, \apjl, 892, L9, \dodoi{10.3847/2041-8213/ab5b9a}

\bibitem[{{Arcavi} {et~al.}(2010){Arcavi}, {Gal-Yam}, {Kasliwal}, {Quimby},
  {Ofek}, {Kulkarni}, {Nugent}, {Cenko}, {Bloom}, {Sullivan}, {Howell},
  {Poznanski}, {Filippenko}, {Law}, {Hook}, {J{\"o}nsson}, {Blake}, {Cooke},
  {Dekany}, {Rahmer}, {Hale}, {Smith}, {Zolkower}, {Velur}, {Walters},
  {Henning}, {Bui}, {McKenna}, \& {Jacobsen}}]{Arcavi2010}
{Arcavi}, I., {Gal-Yam}, A., {Kasliwal}, M.~M., {et~al.} 2010, \apj, 721, 777,
  \dodoi{10.1088/0004-637X/721/1/777}

\bibitem[{{Argast} {et~al.}(2004){Argast}, {Samland}, {Thielemann}, \&
  {Qian}}]{Argast04}
{Argast}, D., {Samland}, M., {Thielemann}, F.~K., \& {Qian}, Y.~Z. 2004, \aap,
  416, 997, \dodoi{10.1051/0004-6361:20034265}

\bibitem[{{Arnould} {et~al.}(2007){Arnould}, {Goriely}, \&
  {Takahashi}}]{Arnould07}
{Arnould}, M., {Goriely}, S., \& {Takahashi}, K. 2007, \physrep, 450, 97,
  \dodoi{10.1016/j.physrep.2007.06.002}

\bibitem[{{Asplund} {et~al.}(2009){Asplund}, {Grevesse}, {Sauval}, \&
  {Scott}}]{Asplund09}
{Asplund}, M., {Grevesse}, N., {Sauval}, A.~J., \& {Scott}, P. 2009, \araa, 47,
  481, \dodoi{10.1146/annurev.astro.46.060407.145222}

\bibitem[{{Barnes} {et~al.}(2018){Barnes}, {Duffell}, {Liu}, {Modjaz},
  {Bianco}, {Kasen}, \& {MacFadyen}}]{Barnes2018}
{Barnes}, J., {Duffell}, P.~C., {Liu}, Y., {et~al.} 2018, \apj, 860, 38,
  \dodoi{10.3847/1538-4357/aabf84}

\bibitem[{{Bauswein} {et~al.}(2013){Bauswein}, {Goriely}, \&
  {Janka}}]{Bauswein13}
{Bauswein}, A., {Goriely}, S., \& {Janka}, H.~T. 2013, \apj, 773, 78,
  \dodoi{10.1088/0004-637X/773/1/78}

\bibitem[{{Bechtol} {et~al.}(2015){Bechtol}, {Drlica-Wagner}, {Balbinot},
  {Pieres}, {Simon}, {Yanny}, {Santiago}, {Wechsler}, {Frieman}, {Walker},
  {Williams}, {Rozo}, {Rykoff}, {Queiroz}, {Luque}, {Benoit-L{\'e}vy},
  {Tucker}, {Sevilla}, {Gruendl}, {da Costa}, {Fausti Neto}, {Maia}, {Abbott},
  {Allam}, {Armstrong}, {Bauer}, {Bernstein}, {Bernstein}, {Bertin}, {Brooks},
  {Buckley-Geer}, {Burke}, {Carnero Rosell}, {Castander}, {Covarrubias},
  {D'Andrea}, {DePoy}, {Desai}, {Diehl}, {Eifler}, {Estrada}, {Evrard},
  {Fernandez}, {Finley}, {Flaugher}, {Gaztanaga}, {Gerdes}, {Girardi},
  {Gladders}, {Gruen}, {Gutierrez}, {Hao}, {Honscheid}, {Jain}, {James},
  {Kent}, {Kron}, {Kuehn}, {Kuropatkin}, {Lahav}, {Li}, {Lin}, {Makler},
  {March}, {Marshall}, {Martini}, {Merritt}, {Miller}, {Miquel}, {Mohr},
  {Neilsen}, {Nichol}, {Nord}, {Ogando}, {Peoples}, {Petravick}, {Plazas},
  {Romer}, {Roodman}, {Sako}, {Sanchez}, {Scarpine}, {Schubnell}, {Smith},
  {Soares-Santos}, {Sobreira}, {Suchyta}, {Swanson}, {Tarle}, {Thaler},
  {Thomas}, {Wester}, {Zuntz}, \& {DES Collaboration}}]{Bechtol15}
{Bechtol}, K., {Drlica-Wagner}, A., {Balbinot}, E., {et~al.} 2015, \apj, 807,
  50, \dodoi{10.1088/0004-637X/807/1/50}

\bibitem[{{Beniamini} {et~al.}(2016){Beniamini}, {Hotokezaka}, \&
  {Piran}}]{Beniamini16}
{Beniamini}, P., {Hotokezaka}, K., \& {Piran}, T. 2016, \apjl, 829, L13,
  \dodoi{10.3847/2041-8205/829/1/L13}

\bibitem[{{Bonetti} {et~al.}(2019){Bonetti}, {Perego}, {Dotti}, \&
  {Cescutti}}]{Bonetti19}
{Bonetti}, M., {Perego}, A., {Dotti}, M., \& {Cescutti}, G. 2019, \mnras, 490,
  296, \dodoi{10.1093/mnras/stz2554}

\bibitem[{{Bramante} \& {Linden}(2016)}]{Bramante16}
{Bramante}, J., \& {Linden}, T. 2016, \apj, 826, 57,
  \dodoi{10.3847/0004-637X/826/1/57}

\bibitem[{{Brauer} {et~al.}(2019){Brauer}, {Ji}, {Frebel}, {Dooley},
  {G{\'o}mez}, \& {O'Shea}}]{Brauer19}
{Brauer}, K., {Ji}, A.~P., {Frebel}, A., {et~al.} 2019, \apj, 871, 247,
  \dodoi{10.3847/1538-4357/aafafb}

\bibitem[{{Bromberg} {et~al.}(2011){Bromberg}, {Nakar}, \&
  {Piran}}]{Bromberg11}
{Bromberg}, O., {Nakar}, E., \& {Piran}, T. 2011, \apjl, 739, L55,
  \dodoi{10.1088/2041-8205/739/2/L55}

\bibitem[{{Burbidge} {et~al.}(1957){Burbidge}, {Burbidge}, {Fowler}, \&
  {Hoyle}}]{Burbidge57}
{Burbidge}, E.~M., {Burbidge}, G.~R., {Fowler}, W.~A., \& {Hoyle}, F. 1957,
  Reviews of Modern Physics, 29, 547, \dodoi{10.1103/RevModPhys.29.547}

\bibitem[{{Cameron}(1957)}]{Cameron57}
{Cameron}, A.~G.~W. 1957, \pasp, 69, 201, \dodoi{10.1086/127051}

\bibitem[{{Cescutti} {et~al.}(2015){Cescutti}, {Romano}, {Matteucci},
  {Chiappini}, \& {Hirschi}}]{Cescutti15}
{Cescutti}, G., {Romano}, D., {Matteucci}, F., {Chiappini}, C., \& {Hirschi},
  R. 2015, \aap, 577, A139, \dodoi{10.1051/0004-6361/201525698}

\bibitem[{{Chaurasia} {et~al.}(2018){Chaurasia}, {Dietrich},
  {Johnson-McDaniel}, {Ujevic}, {Tichy}, \& {Br{\"u}gmann}}]{Chaurasia18}
{Chaurasia}, S.~V., {Dietrich}, T., {Johnson-McDaniel}, N.~K., {et~al.} 2018,
  \prd, 98, 104005, \dodoi{10.1103/PhysRevD.98.104005}

\bibitem[{{Chiaki} \& {Wise}(2019)}]{Chiaki19}
{Chiaki}, G., \& {Wise}, J.~H. 2019, \mnras, 482, 3933,
  \dodoi{10.1093/mnras/sty2984}

\bibitem[{{C{\^o}t{\'e}} {et~al.}(2018){C{\^o}t{\'e}}, {Fryer}, {Belczynski},
  {Korobkin}, {Chru{\'s}li{\'n}ska}, {Vassh}, {Mumpower}, {Lippuner},
  {Sprouse}, {Surman}, \& {Wollaeger}}]{cote18}
{C{\^o}t{\'e}}, B., {Fryer}, C.~L., {Belczynski}, K., {et~al.} 2018, \apj, 855,
  99, \dodoi{10.3847/1538-4357/aaad67}

\bibitem[{{Coulter} {et~al.}(2017){Coulter}, {Foley}, {Kilpatrick}, {Drout},
  {Piro}, {Shappee}, {Siebert}, {Simon}, {Ulloa}, {Kasen}, {Madore},
  {Murguia-Berthier}, {Pan}, {Prochaska}, {Ramirez-Ruiz}, {Rest}, \&
  {Rojas-Bravo}}]{coulter17}
{Coulter}, D.~A., {Foley}, R.~J., {Kilpatrick}, C.~D., {et~al.} 2017, Science,
  358, 1556, \dodoi{10.1126/science.aap9811}

\bibitem[{Davidson-Pilon {et~al.}(2020)Davidson-Pilon, Kalderstam, Jacobson,
  Zivich, Kuhn, Williamson, sean reed, AbdealiJK, Fiore-Gartland, Datta,
  Moneda, Gabriel, WIlson, Parij, Moncada-Torres, Stark, Anton, Peña, Besson,
  Singaravelan, Jona, Gadgil, Golland, Hussey, Kumar, Begun, Noorbakhsh,
  Klintberg, Rendeiro, \& Flaxman}]{lifelines}
Davidson-Pilon, C., Kalderstam, J., Jacobson, N., {et~al.} 2020,
  CamDavidsonPilon/lifelines: v0.24.8, v0.24.8,  Zenodo,
  \dodoi{10.5281/zenodo.3833188}

\bibitem[{{Dekel} \& {Woo}(2003)}]{Dekel03}
{Dekel}, A., \& {Woo}, J. 2003, \mnras, 344, 1131,
  \dodoi{10.1046/j.1365-8711.2003.06923.x}

\bibitem[{{Dietrich} {et~al.}(2015){Dietrich}, {Bernuzzi}, {Ujevic}, \&
  {Br{\"u}gmann}}]{dietrich15}
{Dietrich}, T., {Bernuzzi}, S., {Ujevic}, M., \& {Br{\"u}gmann}, B. 2015, \prd,
  91, 124041, \dodoi{10.1103/PhysRevD.91.124041}

\bibitem[{{Drlica-Wagner} {et~al.}(2015){Drlica-Wagner}, {Bechtol}, {Rykoff},
  {Luque}, {Queiroz}, {Mao}, {Wechsler}, {Simon}, {Santiago}, {Yanny},
  {Balbinot}, {Dodelson}, {Fausti Neto}, {James}, {Li}, {Maia}, {Marshall},
  {Pieres}, {Stringer}, {Walker}, {Abbott}, {Abdalla}, {Allam},
  {Benoit-L{\'e}vy}, {Bernstein}, {Bertin}, {Brooks}, {Buckley-Geer}, {Burke},
  {Carnero Rosell}, {Carrasco Kind}, {Carretero}, {Crocce}, {da Costa},
  {Desai}, {Diehl}, {Dietrich}, {Doel}, {Eifler}, {Evrard}, {Finley},
  {Flaugher}, {Fosalba}, {Frieman}, {Gaztanaga}, {Gerdes}, {Gruen}, {Gruendl},
  {Gutierrez}, {Honscheid}, {Kuehn}, {Kuropatkin}, {Lahav}, {Martini},
  {Miquel}, {Nord}, {Ogando}, {Plazas}, {Reil}, {Roodman}, {Sako}, {Sanchez},
  {Scarpine}, {Schubnell}, {Sevilla-Noarbe}, {Smith}, {Soares-Santos},
  {Sobreira}, {Suchyta}, {Swanson}, {Tarle}, {Tucker}, {Vikram}, {Wester},
  {Zhang}, {Zuntz}, \& {DES Collaboration}}]{Drlica15}
{Drlica-Wagner}, A., {Bechtol}, K., {Rykoff}, E.~S., {et~al.} 2015, \apj, 813,
  109, \dodoi{10.1088/0004-637X/813/2/109}

\bibitem[{{Drout} {et~al.}(2017){Drout}, {Piro}, {Shappee}, {Kilpatrick},
  {Simon}, {Contreras}, {Coulter}, {Foley}, {Siebert}, {Morrell}, {Boutsia},
  {Di Mille}, {Holoien}, {Kasen}, {Kollmeier}, {Madore}, {Monson},
  {Murguia-Berthier}, {Pan}, {Prochaska}, {Ramirez-Ruiz}, {Rest}, {Adams},
  {Alatalo}, {Ba{\~n}ados}, {Baughman}, {Beers}, {Bernstein}, {Bitsakis},
  {Campillay}, {Hansen}, {Higgs}, {Ji}, {Maravelias}, {Marshall}, {Moni Bidin},
  {Prieto}, {Rasmussen}, {Rojas-Bravo}, {Strom}, {Ulloa},
  {Vargas-Gonz{\'a}lez}, {Wan}, \& {Whitten}}]{Drout17}
{Drout}, M.~R., {Piro}, A.~L., {Shappee}, B.~J., {et~al.} 2017, ArXiv e-prints.
\newblock \doarXiv{1710.05443}

\bibitem[{{Dvorkin} {et~al.}(2020){Dvorkin}, {Daigne}, {Goriely}, {Vangioni},
  \& {Silk}}]{Dvorkin20}
{Dvorkin}, I., {Daigne}, F., {Goriely}, S., {Vangioni}, E., \& {Silk}, J. 2020,
  arXiv e-prints, arXiv:2010.00625.
\newblock \doarXiv{2010.00625}

\bibitem[{{Emerick} {et~al.}(2018){Emerick}, {Bryan}, {Mac Low},
  {C{\^o}t{\'e}}, {Johnston}, \& {O'Shea}}]{Emerick18}
{Emerick}, A., {Bryan}, G.~L., {Mac Low}, M.-M., {et~al.} 2018, \apj, 869, 94,
  \dodoi{10.3847/1538-4357/aaec7d}

\bibitem[{{Ezzeddine} {et~al.}(2020){Ezzeddine}, {Rasmussen}, {Frebel},
  {Chiti}, {Hinojisa}, {Placco}, {Roederer}, {Ji}, {Beers}, {Hansen}, {Sakari},
  \& {Melendez}}]{rpa3}
{Ezzeddine}, R., {Rasmussen}, K., {Frebel}, A., {et~al.} 2020, arXiv e-prints,
  arXiv:2006.07731.
\newblock \doarXiv{2006.07731}

\bibitem[{{Feigelson} \& {Nelson}(1985)}]{upperlimits1}
{Feigelson}, E.~D., \& {Nelson}, P.~I. 1985, \apj, 293, 192,
  \dodoi{10.1086/163225}

\bibitem[{{Fong} \& {Berger}(2013)}]{Fong2013}
{Fong}, W., \& {Berger}, E. 2013, \apj, 776, 18,
  \dodoi{10.1088/0004-637X/776/1/18}

\bibitem[{{Fynbo} {et~al.}(2006){Fynbo}, {Watson}, {Th{\"o}ne}, {Sollerman},
  {Bloom}, {Davis}, {Hjorth}, {Jakobsson}, {J{\o}rgensen}, {Graham},
  {Fruchter}, {Bersier}, {Kewley}, {Cassan}, {Castro Cer{\'o}n}, {Foley},
  {Gorosabel}, {Hinse}, {Horne}, {Jensen}, {Klose}, {Kocevski}, {Marquette},
  {Perley}, {Ramirez-Ruiz}, {Stritzinger}, {Vreeswijk}, {Wijers}, {Woller},
  {Xu}, \& {Zub}}]{2006Natur.444.1047F}
{Fynbo}, J. P.~U., {Watson}, D., {Th{\"o}ne}, C.~C., {et~al.} 2006, \nat, 444,
  1047, \dodoi{10.1038/nature05375}

\bibitem[{{Graur} {et~al.}(2017{\natexlab{a}}){Graur}, {Bianco}, {Huang},
  {Modjaz}, {Shivvers}, {Filippenko}, {Li}, \& {Eldridge}}]{Graur2017a}
{Graur}, O., {Bianco}, F.~B., {Huang}, S., {et~al.} 2017{\natexlab{a}}, \apj,
  837, 120, \dodoi{10.3847/1538-4357/aa5eb8}

\bibitem[{{Graur} {et~al.}(2017{\natexlab{b}}){Graur}, {Bianco}, {Modjaz},
  {Shivvers}, {Filippenko}, {Li}, \& {Smith}}]{Graur2017b}
{Graur}, O., {Bianco}, F.~B., {Modjaz}, M., {et~al.} 2017{\natexlab{b}}, \apj,
  837, 121, \dodoi{10.3847/1538-4357/aa5eb7}

\bibitem[{{Griffen} {et~al.}(2018){Griffen}, {Dooley}, {Ji}, {O'Shea},
  {G{\'o}mez}, \& {Frebel}}]{Griffen18}
{Griffen}, B.~F., {Dooley}, G.~A., {Ji}, A.~P., {et~al.} 2018, \mnras, 474,
  443, \dodoi{10.1093/mnras/stx2749}

\bibitem[{{Guetta} {et~al.}(2005){Guetta}, {Piran}, \& {Waxman}}]{Guetta2005}
{Guetta}, D., {Piran}, T., \& {Waxman}, E. 2005, \apj, 619, 412,
  \dodoi{10.1086/423125}

\bibitem[{{Hajela} {et~al.}(2019){Hajela}, {Margutti}, {Alexander},
  {Kathirgamaraju}, {Baldeschi}, {Guidorzi}, {Giannios}, {Fong}, {Wu},
  {MacFadyen}, {Paggi}, {Berger}, {Blanchard}, {Chornock}, {Coppejans},
  {Cowperthwaite}, {Eftekhari}, {Gomez}, {Hosseinzadeh}, {Laskar}, {Metzger},
  {Nicholl}, {Paterson}, {Radice}, {Sironi}, {Terreran}, {Villar}, {Williams},
  {Xie}, \& {Zrake}}]{GW170817Eiso}
{Hajela}, A., {Margutti}, R., {Alexander}, K.~D., {et~al.} 2019, \apjl, 886,
  L17, \dodoi{10.3847/2041-8213/ab5226}

\bibitem[{{Hansen} {et~al.}(2017){Hansen}, {Simon}, {Marshall}, {Li},
  {Carollo}, {DePoy}, {Nagasawa}, {Bernstein}, {Drlica-Wagner}, {Abdalla},
  {Allam}, {Annis}, {Bechtol}, {Benoit-L{\'e}vy}, {Brooks}, {Buckley-Geer},
  {Carnero Rosell}, {Carrasco Kind}, {Carretero}, {Cunha}, {da Costa}, {Desai},
  {Eifler}, {Fausti Neto}, {Flaugher}, {Frieman}, {Garc{\'{\i}}a-Bellido},
  {Gaztanaga}, {Gerdes}, {Gruen}, {Gruendl}, {Gschwend}, {Gutierrez}, {James},
  {Krause}, {Kuehn}, {Kuropatkin}, {Lahav}, {Miquel}, {Plazas}, {Romer},
  {Sanchez}, {Santiago}, {Scarpine}, {Smith}, {Soares-Santos}, {Sobreira},
  {Suchyta}, {Swanson}, {Tarle}, {Walker}, \& {DES Collaboration}}]{Hansen17}
{Hansen}, T.~T., {Simon}, J.~D., {Marshall}, J.~L., {et~al.} 2017, \apj, 838,
  44, \dodoi{10.3847/1538-4357/aa634a}

\bibitem[{{Hansen} {et~al.}(2018){Hansen}, {Holmbeck}, {Beers}, {Placco},
  {Roederer}, {Frebel}, {Sakari}, {Simon}, \& {Thompson}}]{Hansen18}
{Hansen}, T.~T., {Holmbeck}, E.~M., {Beers}, T.~C., {et~al.} 2018, \apj, 858,
  92, \dodoi{10.3847/1538-4357/aabacc}

\bibitem[{{Hansen} {et~al.}(2020){Hansen}, {Marshall}, {Simon}, {Li},
  {Bernstein}, {Pace}, {Ferguson}, {Nagasawa}, {Kuehn}, {Carollo}, {Geha},
  {James}, {Walker}, {Diehl}, {Aguena}, {Allam}, {Avila}, {Bertin}, {Brooks},
  {Buckley-Geer}, {Burke}, {Rosell}, {Kind}, {Carretero}, {Costanzi}, {Da
  Costa}, {Desai}, {De Vicente}, {Doel}, {Eckert}, {Eifler}, {Everett},
  {Ferrero}, {Frieman}, {Garc{\'\i}a-Bellido}, {Gaztanaga}, {Gerdes}, {Gruen},
  {Gruendl}, {Gschwend}, {Gutierrez}, {Hinton}, {Hollowood}, {Honscheid},
  {Kuropatkin}, {Maia}, {March}, {Miquel}, {Palmese}, {Paz-Chinch{\'o}n},
  {Plazas}, {Sanchez}, {Santiago}, {Scarpine}, {Serrano}, {Smith},
  {Soares-Santos}, {Suchyta}, {Swanson}, {Tarle}, {Varga}, {Wilkinson}, \& {DES
  Collaboration}}]{Hansen20_Grus}
{Hansen}, T.~T., {Marshall}, J.~L., {Simon}, J.~D., {et~al.} 2020, \apj, 897,
  183, \dodoi{10.3847/1538-4357/ab9643}

\bibitem[{{Haynes} \& {Kobayashi}(2019)}]{Haynes19}
{Haynes}, C.~J., \& {Kobayashi}, C. 2019, \mnras, 483, 5123,
  \dodoi{10.1093/mnras/sty3389}

\bibitem[{{Holmbeck} {et~al.}(2020){Holmbeck}, {Hansen}, {Beers}, {Placco},
  {Whitten}, {Rasmussen}, {Roederer}, {Ezzeddine}, {Sakari}, {Frebel}, {Drout},
  {Simon}, {Thompson}, {Bland-Hawthorn}, {Gibson}, {Grebel}, {Kordopatis},
  {Kunder}, {Mel{\'e}ndez}, {Navarro}, {Reid}, {Seabroke}, {Steinmetz},
  {Watson}, \& {Wyse}}]{RPA4}
{Holmbeck}, E.~M., {Hansen}, T.~T., {Beers}, T.~C., {et~al.} 2020, \apjs, 249,
  30, \dodoi{10.3847/1538-4365/ab9c19}

\bibitem[{{Hotokezaka} {et~al.}(2013){Hotokezaka}, {Kiuchi}, {Kyutoku},
  {Okawa}, {Sekiguchi}, {Shibata}, \& {Taniguchi}}]{Hotokezaka13}
{Hotokezaka}, K., {Kiuchi}, K., {Kyutoku}, K., {et~al.} 2013, \prd, 87, 024001,
  \dodoi{10.1103/PhysRevD.87.024001}

\bibitem[{{Ishimaru} {et~al.}(2015){Ishimaru}, {Wanajo}, \&
  {Prantzos}}]{Ishimaru15}
{Ishimaru}, Y., {Wanajo}, S., \& {Prantzos}, N. 2015, \apjl, 804, L35,
  \dodoi{10.1088/2041-8205/804/2/L35}

\bibitem[{{Ji} {et~al.}(2015{\natexlab{a}}){Ji}, {Frebel}, \&
  {Bromm}}]{2015MNRAS.454..659J}
{Ji}, A.~P., {Frebel}, A., \& {Bromm}, V. 2015{\natexlab{a}}, \mnras, 454, 659,
  \dodoi{10.1093/mnras/stv2052}

\bibitem[{{Ji} {et~al.}(2015{\natexlab{b}}){Ji}, {Frebel}, \& {Bromm}}]{Ji15}
---. 2015{\natexlab{b}}, \mnras, 454, 659, \dodoi{10.1093/mnras/stv2052}

\bibitem[{{Ji} {et~al.}(2016){Ji}, {Frebel}, {Chiti}, \& {Simon}}]{Ji16b}
{Ji}, A.~P., {Frebel}, A., {Chiti}, A., \& {Simon}, J.~D. 2016, \nat, 531, 610,
  \dodoi{10.1038/nature17425}

\bibitem[{{Kirby} {et~al.}(2011){Kirby}, {Martin}, \&
  {Finlator}}]{Kirby11outflow}
{Kirby}, E.~N., {Martin}, C.~L., \& {Finlator}, K. 2011, \apjl, 742, L25,
  \dodoi{10.1088/2041-8205/742/2/L25}

\bibitem[{{Kobayashi} {et~al.}(2020){Kobayashi}, {Karakas}, \&
  {Lugaro}}]{Kobayashi20}
{Kobayashi}, C., {Karakas}, A.~I., \& {Lugaro}, M. 2020, arXiv e-prints,
  arXiv:2008.04660.
\newblock \doarXiv{2008.04660}

\bibitem[{{Korobkin} {et~al.}(2012){Korobkin}, {Rosswog}, {Arcones}, \&
  {Winteler}}]{Korobkin12}
{Korobkin}, O., {Rosswog}, S., {Arcones}, A., \& {Winteler}, C. 2012, \mnras,
  426, 1940, \dodoi{10.1111/j.1365-2966.2012.21859.x}

\bibitem[{{Lazzati} {et~al.}(2012){Lazzati}, {Morsony}, {Blackwell}, \&
  {Begelman}}]{Lazzati2012}
{Lazzati}, D., {Morsony}, B.~J., {Blackwell}, C.~H., \& {Begelman}, M.~C. 2012,
  \apj, 750, 68, \dodoi{10.1088/0004-637X/750/1/68}

\bibitem[{{Li} {et~al.}(2011{\natexlab{a}}){Li}, {Chornock}, {Leaman},
  {Filippenko}, {Poznanski}, {Wang}, {Ganeshalingam}, \& {Mannucci}}]{Li2011}
{Li}, W., {Chornock}, R., {Leaman}, J., {et~al.} 2011{\natexlab{a}}, \mnras,
  412, 1473, \dodoi{10.1111/j.1365-2966.2011.18162.x}

\bibitem[{{Li} {et~al.}(2011{\natexlab{b}}){Li}, {Leaman}, {Chornock},
  {Filippenko}, {Poznanski}, {Ganeshalingam}, {Wang}, {Modjaz}, {Jha}, {Foley},
  \& {Smith}}]{LOSS11}
{Li}, W., {Leaman}, J., {Chornock}, R., {et~al.} 2011{\natexlab{b}}, \mnras,
  412, 1441, \dodoi{10.1111/j.1365-2966.2011.18160.x}

\bibitem[{{Lien} {et~al.}(2014){Lien}, {Sakamoto}, {Gehrels}, {Palmer},
  {Barthelmy}, {Graziani}, \& {Cannizzo}}]{Lien2014}
{Lien}, A., {Sakamoto}, T., {Gehrels}, N., {et~al.} 2014, \apj, 783, 24,
  \dodoi{10.1088/0004-637X/783/1/24}

\bibitem[{{MacFadyen} \& {Woosley}(1999)}]{MacFadyen99}
{MacFadyen}, A.~I., \& {Woosley}, S.~E. 1999, \apj, 524, 262,
  \dodoi{10.1086/307790}

\bibitem[{{Macias} \& {Ramirez-Ruiz}(2018)}]{Macias18}
{Macias}, P., \& {Ramirez-Ruiz}, E. 2018, \apj, 860, 89,
  \dodoi{10.3847/1538-4357/aac3e0}

\bibitem[{{Macias} \& {Ramirez-Ruiz}(2019)}]{Macias19}
---. 2019, \apjl, 877, L24, \dodoi{10.3847/2041-8213/ab2049}

\bibitem[{{Magg} {et~al.}(2020){Magg}, {Nordlander}, {Glover}, {Hansen},
  {Ishigaki}, {Heger}, {Klessen}, {Kobayashi}, \& {Nomoto}}]{Magg20}
{Magg}, M., {Nordlander}, T., {Glover}, S. C.~O., {et~al.} 2020, \mnras, 498,
  3703, \dodoi{10.1093/mnras/staa2624}

\bibitem[{{Margalit} \& {Metzger}(2017)}]{Margalit17}
{Margalit}, B., \& {Metzger}, B.~D. 2017, \apjl, 850, L19,
  \dodoi{10.3847/2041-8213/aa991c}

\bibitem[{{Margutti} {et~al.}(2014){Margutti}, {Milisavljevic}, {Soderberg},
  {Guidorzi}, {Morsony}, {Sanders}, {Chakraborti}, {Ray}, {Kamble}, {Drout},
  {Parrent}, {Zauderer}, \& {Chomiuk}}]{Margutti2014}
{Margutti}, R., {Milisavljevic}, D., {Soderberg}, A.~M., {et~al.} 2014, \apj,
  797, 107, \dodoi{10.1088/0004-637X/797/2/107}

\bibitem[{{McLaughlin} \& {Surman}(2005)}]{McLaughlin2005}
{McLaughlin}, G.~C., \& {Surman}, R. 2005, \nphysa, 758, 189,
  \dodoi{10.1016/j.nuclphysa.2005.05.036}

\bibitem[{{McQuinn} {et~al.}(2015){McQuinn}, {Skillman}, {Dolphin}, {Cannon},
  {Salzer}, {Rhode}, {Adams}, {Berg}, {Giovanelli}, \& {Haynes}}]{McQuinn15}
{McQuinn}, K. B.~W., {Skillman}, E.~D., {Dolphin}, A., {et~al.} 2015, \apjl,
  815, L17, \dodoi{10.1088/2041-8205/815/2/L17}

\bibitem[{{Milisavljevic} {et~al.}(2015){Milisavljevic}, {Margutti}, {Parrent},
  {Soderberg}, {Fesen}, {Mazzali}, {Maeda}, {Sanders}, {Cenko}, {Silverman},
  {Filippenko}, {Kamble}, {Chakraborti}, {Drout}, {Kirshner}, {Pickering},
  {Kawabata}, {Hattori}, {Hsiao}, {Stritzinger}, {Marion}, {Vinko}, \&
  {Wheeler}}]{Milisavljevic2015}
{Milisavljevic}, D., {Margutti}, R., {Parrent}, J.~T., {et~al.} 2015, \apj,
  799, 51, \dodoi{10.1088/0004-637X/799/1/51}

\bibitem[{{Miller} {et~al.}(2019){Miller}, {Sprouse}, {Fryer}, {Ryan},
  {Dolence}, {Mumpower}, \& {Surman}}]{Miller19}
{Miller}, J.~M., {Sprouse}, T.~M., {Fryer}, C.~L., {et~al.} 2019, arXiv
  e-prints, arXiv:1912.03378.
\newblock \doarXiv{1912.03378}

\bibitem[{{Modjaz} {et~al.}(2016){Modjaz}, {Liu}, {Bianco}, \&
  {Graur}}]{Modjaz2016}
{Modjaz}, M., {Liu}, Y.~Q., {Bianco}, F.~B., \& {Graur}, O. 2016, \apj, 832,
  108, \dodoi{10.3847/0004-637X/832/2/108}

\bibitem[{{Modjaz} {et~al.}(2020){Modjaz}, {Bianco}, {Siwek}, {Huang},
  {Perley}, {Fierroz}, {Liu}, {Arcavi}, {Gal-Yam}, {Filippenko},
  {Blagorodnova}, {Cenko}, {Kasliwal}, {Kulkarni}, {Schulze}, {Taggart}, \&
  {Zheng}}]{Modjaz2020}
{Modjaz}, M., {Bianco}, F.~B., {Siwek}, M., {et~al.} 2020, \apj, 892, 153,
  \dodoi{10.3847/1538-4357/ab4185}

\bibitem[{{M{\"o}sta} {et~al.}(2018){M{\"o}sta}, {Roberts}, {Halevi}, {Ott},
  {Lippuner}, {Haas}, \& {Schnetter}}]{Mosta18}
{M{\"o}sta}, P., {Roberts}, L.~F., {Halevi}, G., {et~al.} 2018, \apj, 864, 171,
  \dodoi{10.3847/1538-4357/aad6ec}

\bibitem[{{Mu{\~n}oz} {et~al.}(2018){Mu{\~n}oz}, {C{\^o}t{\'e}}, {Santana},
  {Geha}, {Simon}, {Oyarz{\'u}n}, {Stetson}, \& {Djorgovski}}]{Munoz18}
{Mu{\~n}oz}, R.~R., {C{\^o}t{\'e}}, P., {Santana}, F.~A., {et~al.} 2018, \apj,
  860, 66, \dodoi{10.3847/1538-4357/aac16b}

\bibitem[{{Mutlu-Pakdil} {et~al.}(2018){Mutlu-Pakdil}, {Sand}, {Carlin},
  {Spekkens}, {Caldwell}, {Crnojevi{\'c}}, {Hughes}, {Willman}, \&
  {Zaritsky}}]{Mutlu18}
{Mutlu-Pakdil}, B., {Sand}, D.~J., {Carlin}, J.~L., {et~al.} 2018, \apj, 863,
  25, \dodoi{10.3847/1538-4357/aacd0e}

\bibitem[{{Nishimura} {et~al.}(2015){Nishimura}, {Takiwaki}, \&
  {Thielemann}}]{Nishimura15}
{Nishimura}, N., {Takiwaki}, T., \& {Thielemann}, F.-K. 2015, \apj, 810, 109,
  \dodoi{10.1088/0004-637X/810/2/109}

\bibitem[{{Ojima} {et~al.}(2018){Ojima}, {Ishimaru}, {Wanajo}, {Prantzos}, \&
  {Francois}}]{Ojima18}
{Ojima}, T., {Ishimaru}, Y., {Wanajo}, S., {Prantzos}, N., \& {Francois}, P.
  2018, ArXiv e-prints, arXiv:1808.03390.
\newblock \doarXiv{1808.03390}

\bibitem[{{Papenfort} {et~al.}(2018){Papenfort}, {Gold}, \&
  {Rezzolla}}]{Papenfort18}
{Papenfort}, L.~J., {Gold}, R., \& {Rezzolla}, L. 2018, \prd, 98, 104028,
  \dodoi{10.1103/PhysRevD.98.104028}

\bibitem[{{Petropoulou} {et~al.}(2017){Petropoulou}, {Barniol Duran}, \&
  {Giannios}}]{petropoulou17}
{Petropoulou}, M., {Barniol Duran}, R., \& {Giannios}, D. 2017, \mnras, 472,
  2722, \dodoi{10.1093/mnras/stx2151}

\bibitem[{{Pian} {et~al.}(2017){Pian}, {D'Avanzo}, {Benetti}, {Branchesi},
  {Brocato}, {Campana}, {Cappellaro}, {Covino}, {D'Elia}, {Fynbo}, {Getman},
  {Ghirland a}, {Ghisellini}, {Grado}, {Greco}, {Hjorth}, {Kouveliotou},
  {Levan}, {Limatola}, {Malesani}, {Mazzali}, {Melandri}, {M{\o}ller},
  {Nicastro}, {Palazzi}, {Piranomonte}, {Rossi}, {Salafia}, {Selsing},
  {Stratta}, {Tanaka}, {Tanvir}, {Tomasella}, {Watson}, {Yang}, {Amati},
  {Antonelli}, {Ascenzi}, {Bernardini}, {Bo{\"e}r}, {Bufano}, {Bulgarelli},
  {Capaccioli}, {Casella}, {Castro-Tirado}, {Chassande-Mottin}, {Ciolfi},
  {Copperwheat}, {Dadina}, {De Cesare}, {di Paola}, {Fan}, {Gendre},
  {Giuffrida}, {Giunta}, {Hunt}, {Israel}, {Jin}, {Kasliwal}, {Klose}, {Lisi},
  {Longo}, {Maiorano}, {Mapelli}, {Masetti}, {Nava}, {Patricelli}, {Perley},
  {Pescalli}, {Piran}, {Possenti}, {Pulone}, {Razzano}, {Salvaterra},
  {Schipani}, {Spera}, {Stamerra}, {Stella}, {Tagliaferri}, {Testa}, {Troja},
  {Turatto}, {Vergani}, \& {Vergani}}]{Pian17}
{Pian}, E., {D'Avanzo}, P., {Benetti}, S., {et~al.} 2017, \nat, 551, 67,
  \dodoi{10.1038/nature24298}

\bibitem[{{Radice} {et~al.}(2018){Radice}, {Perego}, {Hotokezaka}, {Fromm},
  {Bernuzzi}, \& {Roberts}}]{Radice18}
{Radice}, D., {Perego}, A., {Hotokezaka}, K., {et~al.} 2018, \apj, 869, 130,
  \dodoi{10.3847/1538-4357/aaf054}

\bibitem[{{Ramirez-Ruiz} {et~al.}(2015){Ramirez-Ruiz}, {Trenti}, {MacLeod},
  {Roberts}, {Lee}, \& {Saladino-Rosas}}]{RamirezRuiz15}
{Ramirez-Ruiz}, E., {Trenti}, M., {MacLeod}, M., {et~al.} 2015, \apjl, 802,
  L22, \dodoi{10.1088/2041-8205/802/2/L22}

\bibitem[{{Robertson} {et~al.}(2005){Robertson}, {Bullock}, {Font}, {Johnston},
  \& {Hernquist}}]{Robertson05}
{Robertson}, B., {Bullock}, J.~S., {Font}, A.~S., {Johnston}, K.~V., \&
  {Hernquist}, L. 2005, \apj, 632, 872, \dodoi{10.1086/452619}

\bibitem[{{Roederer} {et~al.}(2014{\natexlab{a}}){Roederer}, {Cowan},
  {Preston}, {Shectman}, {Sneden}, \& {Thompson}}]{2014MNRAS.445.2946R}
{Roederer}, I.~U., {Cowan}, J.~J., {Preston}, G.~W., {et~al.}
  2014{\natexlab{a}}, \mnras, 445, 2970, \dodoi{10.1093/mnras/stu1977}

\bibitem[{{Roederer} {et~al.}(2014{\natexlab{b}}){Roederer}, {Preston},
  {Thompson}, {Shectman}, \& {Sneden}}]{Roederer14b}
{Roederer}, I.~U., {Preston}, G.~W., {Thompson}, I.~B., {Shectman}, S.~A., \&
  {Sneden}, C. 2014{\natexlab{b}}, \apj, 784, 158,
  \dodoi{10.1088/0004-637X/784/2/158}

\bibitem[{{Safarzadeh} {et~al.}(2019{\natexlab{a}}){Safarzadeh},
  {Ramirez-Ruiz}, {Andrews}, {Macias}, {Fragos}, \&
  {Scannapieco}}]{Safarzadeh19b}
{Safarzadeh}, M., {Ramirez-Ruiz}, E., {Andrews}, J.~J., {et~al.}
  2019{\natexlab{a}}, \apj, 872, 105, \dodoi{10.3847/1538-4357/aafe0e}

\bibitem[{{Safarzadeh} {et~al.}(2019{\natexlab{b}}){Safarzadeh}, {Sarmento}, \&
  {Scannapieco}}]{Safarzadeh19}
{Safarzadeh}, M., {Sarmento}, R., \& {Scannapieco}, E. 2019{\natexlab{b}},
  \apj, 876, 28, \dodoi{10.3847/1538-4357/ab1341}

\bibitem[{{Sakari} {et~al.}(2018){Sakari}, {Placco}, {Farrell}, {Roederer},
  {Wallerstein}, {Beers}, {Ezzeddine}, {Frebel}, {Hansen}, {Holmbeck},
  {Sneden}, {Cowan}, {Venn}, {Davis}, {Matijevi{\v{c}}}, {Wyse},
  {Bland-Hawthorn}, {Chiappini}, {Freeman}, {Gibson}, {Grebel}, {Helmi},
  {Kordopatis}, {Kunder}, {Navarro}, {Reid}, {Seabroke}, {Steinmetz}, \&
  {Watson}}]{Sakari18}
{Sakari}, C.~M., {Placco}, V.~M., {Farrell}, E.~M., {et~al.} 2018, \apj, 868,
  110, \dodoi{10.3847/1538-4357/aae9df}

\bibitem[{{Schmitt}(1985)}]{upperlimits2}
{Schmitt}, J.~H.~M.~M. 1985, \apj, 293, 178, \dodoi{10.1086/163224}

\bibitem[{{Schulze} {et~al.}(2020){Schulze}, {Yaron}, {Sollerman}, {Leloudas},
  {Gal}, {Wright}, {Lunnan}, {Gal-Yam}, {Ofek}, {Perley}, {Filippenko},
  {Kasliwal}, {Kulkarni}, {Nugent}, {Quimby}, {Sullivan}, {Linn Strothjohann},
  {Arcavi}, {Ben-Ami}, {Bianco}, {Bloom}, {De}, {Fraser}, {Fremling}, {Horesh},
  {Johansson}, {Kelly}, {Knezevic}, {Maguire}, {Nyholm}, {Semeli
  Papadogiannakis}, {Petrushevska}, {Rubin}, {Yan}, {Yang}, {Adams}, {Bufano},
  {Clubb}, {Foley}, {Green}, {Harmanen}, {Ho}, {Hook}, {Hosseinzadeh},
  {Howell}, {Kong}, {Kotak}, {Matheson}, {McCully}, {Milisavljevic}, {Pan},
  {Poznanski}, {Shivvers}, \& {van Velzen}}]{2020arXiv200805988S}
{Schulze}, S., {Yaron}, O., {Sollerman}, J., {et~al.} 2020, arXiv e-prints,
  arXiv:2008.05988.
\newblock \doarXiv{2008.05988}

\bibitem[{{Sekiguchi} {et~al.}(2016){Sekiguchi}, {Kiuchi}, {Kyutoku},
  {Shibata}, \& {Taniguchi}}]{Sekiguchi16}
{Sekiguchi}, Y., {Kiuchi}, K., {Kyutoku}, K., {Shibata}, M., \& {Taniguchi}, K.
  2016, \prd, 93, 124046, \dodoi{10.1103/PhysRevD.93.124046}

\bibitem[{{Shen} {et~al.}(2015){Shen}, {Cooke}, {Ramirez-Ruiz}, {Madau},
  {Mayer}, \& {Guedes}}]{Shen15}
{Shen}, S., {Cooke}, R.~J., {Ramirez-Ruiz}, E., {et~al.} 2015, \apj, 807, 115,
  \dodoi{10.1088/0004-637X/807/2/115}

\bibitem[{{Shivvers} {et~al.}(2017){Shivvers}, {Modjaz}, {Zheng}, {Liu},
  {Filippenko}, {Silverman}, {Matheson}, {Pastorello}, {Graur}, {Foley},
  {Chornock}, {Smith}, {Leaman}, \& {Benetti}}]{2017PASP..129e4201S}
{Shivvers}, I., {Modjaz}, M., {Zheng}, W., {et~al.} 2017, \pasp, 129, 054201,
  \dodoi{10.1088/1538-3873/aa54a6}

\bibitem[{{Siegel}(2019)}]{siegel2019b}
{Siegel}, D.~M. 2019, European Physical Journal A, 55, 203,
  \dodoi{10.1140/epja/i2019-12888-9}

\bibitem[{{Siegel}(2020)}]{Siegel20}
---. 2020, arXiv e-prints, arXiv:2008.06078.
\newblock \doarXiv{2008.06078}

\bibitem[{{Siegel} {et~al.}(2019){Siegel}, {Barnes}, \& {Metzger}}]{siegel19}
{Siegel}, D.~M., {Barnes}, J., \& {Metzger}, B.~D. 2019, \nat, 569, 241,
  \dodoi{10.1038/s41586-019-1136-0}

\bibitem[{{Simcoe} {et~al.}(2004){Simcoe}, {Sargent}, \& {Rauch}}]{Simcoe04}
{Simcoe}, R.~A., {Sargent}, W. L.~W., \& {Rauch}, M. 2004, \apj, 606, 92,
  \dodoi{10.1086/382777}

\bibitem[{{Sk{\'u}lad{\'o}ttir} {et~al.}(2019){Sk{\'u}lad{\'o}ttir}, {Hansen},
  {Salvadori}, \& {Choplin}}]{Skuladottir19}
{Sk{\'u}lad{\'o}ttir}, {\'A}., {Hansen}, C.~J., {Salvadori}, S., \& {Choplin},
  A. 2019, \aap, 631, A171, \dodoi{10.1051/0004-6361/201936125}

\bibitem[{{Sneden} {et~al.}(2008){Sneden}, {Cowan}, \& {Gallino}}]{Sneden08}
{Sneden}, C., {Cowan}, J.~J., \& {Gallino}, R. 2008, \araa, 46, 241,
  \dodoi{10.1146/annurev.astro.46.060407.145207}

\bibitem[{{Sobacchi} {et~al.}(2017){Sobacchi}, {Granot}, {Bromberg}, \&
  {Sormani}}]{sobacchi17}
{Sobacchi}, E., {Granot}, J., {Bromberg}, O., \& {Sormani}, M.~C. 2017, \mnras,
  472, 616, \dodoi{10.1093/mnras/stx2083}

\bibitem[{{Soderberg} {et~al.}(2010){Soderberg}, {Chakraborti}, {Pignata},
  {Chevalier}, {Chandra}, {Ray}, {Wieringa}, {Copete}, {Chaplin},
  {Connaughton}, {Barthelmy}, {Bietenholz}, {Chugai}, {Stritzinger}, {Hamuy},
  {Fransson}, {Fox}, {Levesque}, {Grindlay}, {Challis}, {Foley}, {Kirshner},
  {Milne}, \& {Torres}}]{Soderberg2010}
{Soderberg}, A.~M., {Chakraborti}, S., {Pignata}, G., {et~al.} 2010, \nat, 463,
  513, \dodoi{10.1038/nature08714}

\bibitem[{{Surman} {et~al.}(2006){Surman}, {McLaughlin}, \& {Hix}}]{Surman06}
{Surman}, R., {McLaughlin}, G.~C., \& {Hix}, W.~R. 2006, \apj, 643, 1057,
  \dodoi{10.1086/501116}

\bibitem[{{Tarumi} {et~al.}(2020){Tarumi}, {Yoshida}, \& {Inoue}}]{Tarumi20}
{Tarumi}, Y., {Yoshida}, N., \& {Inoue}, S. 2020, \mnras, 494, 120,
  \dodoi{10.1093/mnras/staa720}

\bibitem[{{The LIGO Scientific Collaboration} {et~al.}(2020){The LIGO
  Scientific Collaboration}, {the Virgo Collaboration}, {Abbott}, {Abbott},
  {Abraham}, {Acernese}, {Ackley}, {Adams}, {Adams}, {Adhikari}, {Adya},
  {Affeldt}, {Agathos}, {Agatsuma}, {Aggarwal}, {Aguiar}, {Aiello}, {Ain},
  {Ajith}, {Allen}, {Allocca}, {Altin}, {Amato}, {Anand}, {Ananyeva},
  {Anderson}, {Anderson}, {Angelova}, {Ansoldi}, {Antelis}, {Antier}, {Appert},
  {Arai}, {Araya}, {Areeda}, {Ar{\`e}ne}, {Arnaud}, {Aronson}, {Arun}, {Asali},
  {Ascenzi}, {Ashton}, {Aston}, {Astone}, {Aubin}, {Aufmuth}, {AultONeal},
  {Austin}, {Avendano}, {Babak}, {Badaracco}, {Bader}, {Bae}, {Baer},
  {Bagnasco}, {Baird}, {Ball}, {Ballardin}, {Ballmer}, {Bals}, {Balsamo},
  {Baltus}, {Banagiri}, {Bankar}, {Bankar}, {Barayoga}, {Barbieri}, {Barish},
  {Barker}, {Barneo}, {Barnum}, {Barone}, {Barr}, {Barsotti}, {Barsuglia},
  {Barta}, {Bartlett}, {Bartos}, {Bassiri}, {Basti}, {Bawaj}, {Bayley},
  {Bazzan}, {Becher}, {B{\'e}csy}, {Bedakihale}, {Bejger}, {Belahcene},
  {Beniwal}, {Benjamin}, {Bennett}, {Bentley}, {Bergamin}, {Berger},
  {Bergmann}, {Bernuzzi}, {Berry}, {Bersanetti}, {Bertolini}, {Betzwieser},
  {Bhandare}, {Bhandari}, {Bhattacharjee}, {Bidler}, {Bilenko}, {Billingsley},
  {Birney}, {Birnholtz}, {Biscans}, {Bischi}, {Biscoveanu}, {Bisht}, {Bitossi},
  {Bizouard}, {Blackburn}, {Blackman}, {Blair}, {Blair}, {Blair}, {Blanch},
  {Bobba}, {Bode}, {Boer}, {Boetzel}, {Bogaert}, {Boldrini}, {Bondu},
  {Bonilla}, {Bonnand}, {Booker}, {Boom}, {Bork}, {Boschi}, {Bose},
  {Bossilkov}, {Boudart}, {Bouffanais}, {Bozzi}, {Bradaschia}, {Brady},
  {Bramley}, {Branchesi}, {Brau}, {Breschi}, {Briant}, {Briggs}, {Brighenti},
  {Brillet}, {Brinkmann}, {Brockill}, {Brooks}, {Brooks}, {Brown}, {Brunett},
  {Bruno}, {Bruntz}, {Buikema}, {Bulik}, {Bulten}, {Buonanno}, {Buscicchio},
  {Buskulic}, {Byer}, {Cabero}, {Cadonati}, {Caesar}, {Cagnoli}, {Cahillane},
  {Calder{\'o}n Bustillo}, {Callaghan}, {Callister}, {Calloni}, {Camp},
  {Canepa}, {Cannon}, {Cao}, {Cao}, {Carapella}, {Carbognani}, {Carney},
  {Carpinelli}, {Carullo}, {Carver}, {Casanueva Diaz}, {Casentini}, {Caudill},
  {Cavagli{\`a}}, {Cavalier}, {Cavalieri}, {Cella}, {Cerd{\'a}-Dur{\'a}n},
  {Cesarini}, {Chaibi}, {Chakravarti}, {Chan}, {Chan}, {Chandra}, {Chanial},
  {Chao}, {Charlton}, {Chase}, {Chassande-Mottin}, {Chatterjee},
  {Chattopadhyay}, {Chaturvedi}, {Chatziioannou}, {Chen}, {Chen}, {Chen},
  {Chen}, {Cheng}, {Cheong}, {Chia}, {Chiadini}, {Chierici}, {Chincarini},
  {Chiummo}, {Cho}, {Cho}, {Cho}, {Choate}, {Christensen}, {Chu}, {Chua},
  {Chung}, {Chung}, {Ciani}, {Ciecielag}, {Cie{\'s}lar}, {Cifaldi}, {Ciobanu},
  {Ciolfi}, {Cipriano}, {Cirone}, {Clara}, {Clark}, {Clark}, {Clarke},
  {Clearwater}, {Clesse}, {Cleva}, {Coccia}, {Cohadon}, {Cohen}, {Colleoni},
  {Collette}, {Collins}, {Colpi}, {Constancio}, {Conti}, {Cooper}, {Corban},
  {Corbitt}, {Cordero-Carri{\'o}n}, {Corezzi}, {Corley}, {Cornish}, {Corre},
  {Corsi}, {Cortese}, {Costa}, {Cotesta}, {Coughlin}, {Coughlin}, {Coulon},
  {Countryman}, {Couvares}, {Covas}, {Coward}, {Cowart}, {Coyne}, {Coyne},
  {Creighton}, {Creighton}, {Croquette}, {Crowder}, {Cudell}, {Cullen},
  {Cumming}, {Cummings}, {Cunningham}, {Cuoco}, {Cury\{l\}o}, {Dal Canton},
  {D{\'a}lya}, {Dana}, {DaneshgaranBajastani}, {D'Angelo}, {Danilishin},
  {D'Antonio}, {Danzmann}, {Darsow-Fromm}, {Dasgupta}, {Datrier}, {Dattilo},
  {Dave}, {Davier}, {Davies}, {Davis}, {Daw}, {Dean}, {DeBra}, {Deenadayalan},
  {Degallaix}, {De Laurentis}, {Del{\'e}glise}, {Del Favero}, {De Lillo}, {De
  Lillo}, {Del Pozzo}, {DeMarchi}, {De Matteis}, {D'Emilio}, {Demos}, {Denker},
  {Dent}, {Depasse}, {De Pietri}, {De Rosa}, {De Rossi}, {DeSalvo}, {de
  Varona}, {Dhurandhar}, {D{\'\i}az}, {Diaz-Ortiz}, {Didio}, {Dietrich}, {Di
  Fiore}, {DiFronzo}, {Di Giorgio}, {Di Giovanni}, {Di Giovanni}, {Di
  Girolamo}, {Di Lieto}, {Ding}, {Di Pace}, {Di Palma}, {Di Renzo},
  {Divakarla}, {Dmitriev}, {Doctor}, {D'Onofrio}, {Donovan}, {Dooley},
  {Doravari}, {Dorrington}, {Downes}, {Drago}, {Driggers}, {Du}, {Ducoin},
  {Dupej}, {Durante}, {D'Urso}, {Duverne}, {Dwyer}, {Easter}, {Eddolls},
  {Edelman}, {Edo}, {Edy}, {Effler}, {Eichholz}, {Eikenberry}, {Eisenmann},
  {Eisenstein}, {Ejlli}, {Errico}, {Essick}, {Estell{\'e}s}, {Estevez},
  {Etienne}, {Etzel}, {Evans}, {Evans}, {Ewing}, {Fafone}, {Fair}, {Fairhurst},
  {Fan}, {Farah}, {Farinon}, {Farr}, {Farr}, {Fauchon-Jones}, {Favata}, {Fays},
  {Fazio}, {Feicht}, {Fejer}, {Feng}, {Fenyvesi}, {Ferguson},
  {Fernandez-Galiana}, {Ferrante}, {Ferreira}, {Fidecaro}, {Figura}, {Fiori},
  {Fiorucci}, {Fishbach}, {Fisher}, {Fishner}, {Fittipaldi}, {Fitz-Axen},
  {Fiumara}, {Flaminio}, {Floden}, {Flynn}, {Fong}, {Font}, {Forsyth},
  {Fournier}, {Frasca}, {Frasconi}, {Frei}, {Freise}, {Frey}, {Frey},
  {Fritschel}, {Frolov}, {Fronz{\'e}}, {Fulda}, {Fyffe}, {Gabbard}, {Gadre},
  {Gaebel}, {Gair}, {Gais}, {Galaudage}, {Gamba}, {Ganapathy}, {Ganguly},
  {Gaonkar}, {Garaventa}, {Garc{\'\i}a-Quir{\'o}s}, {Garufi}, {Gateley},
  {Gaudio}, {Gayathri}, {Gemme}, {Gennai}, {George}, {George}, {Gergely},
  {Ghonge}, {Ghosh}, {Ghosh}, {Ghosh}, {Giacomazzo}, {Giacoppo}, {Giaime},
  {Giardina}, {Gibson}, {Gier}, {Gill}, {Giri}, {Glanzer}, {Gleckl}, {Godwin},
  {Goetz}, {Goetz}, {Gohlke}, {Goncharov}, {Gonz{\'a}lez}, {Gopakumar},
  {Gossan}, {Gosselin}, {Gouaty}, {Grace}, {Grado}, {Granata}, {Granata},
  {Grant}, {Gras}, {Grassia}, {Gray}, {Gray}, {Greco}, {Green}, {Green},
  {Gretarsson}, {Griggs}, {Grignani}, {Grimaldi}, {Grimes}, {Grimm}, {Grote},
  {Grunewald}, {Gruning}, {Guerrero}, {Guidi}, {Guimaraes}, {Guix{\'e}},
  {Gulati}, {Guo}, {Gupta}, {Gupta}, {Gupta}, {Gustafson}, {Gustafson},
  {Guzman}, {Haegel}, {Halim}, {Hall}, {Hamilton}, {Hammond}, {Haney}, {Hanke},
  {Hanks}, {Hanna}, {Hannuksela}, {Hansen}, {Hansen}, {Hanson}, {Harder},
  {Hardwick}, {Haris}, {Harms}, {Harry}, {Harry}, {Hartwig}, {Hasskew},
  {Haster}, {Haughian}, {Hayes}, {Healy}, {Heidmann}, {Heintze}, {Heinze},
  {Heinzel}, {Heitmann}, {Hellman}, {Hello}, {Helmling-Cornell}, {Hemming},
  {Hendry}, {Heng}, {Hennes}, {Hennig}, {Hennig}, {Hernandez Vivanco}, {Heurs},
  {Hild}, {Hill}, {Hines}, {Hochheim}, {Hofgard}, {Hofman}, {Hohmann},
  {Holgado}, {Holland}, {Hollows}, {Holmes}, {Holt}, {Holz}, {Hopkins},
  {Horst}, {Hough}, {Howell}, {Hoy}, {Hoyland}, {Huang}, {H{\"u}bner},
  {Huddart}, {Huerta}, {Hughey}, {Hui}, {Husa}, {Huttner}, {Hutzler},
  {Huxford}, {Huynh-Dinh}, {Idzkowski}, {Iess}, {Imperato}, {Inchauspe},
  {Ingram}, {Intini}, {Isi}, {Iyer}, {JaberianHamedan}, {Jacqmin}, {Jadhav},
  {Jadhav}, {James}, {Jani}, {Janssens}, {Janthalur}, {Jaranowski}, {Jariwala},
  {Jaume}, {Jenkins}, {Jeunon}, {Jiang}, {Johns}, {Jones}, {Jones}, {Jones},
  {Jones}, {Jones}, {Jonker}, {Ju}, {Junker}, {Kalaghatgi}, {Kalogera},
  {Kamai}, {Kandhasamy}, {Kang}, {Kanner}, {Kapadia}, {Kapasi}, {Karathanasis},
  {Karki}, {Kashyap}, {Kasprzack}, {Kastaun}, {Katsanevas}, {Katsavounidis},
  {Katzman}, {Kawabe}, {K{\'e}f{\'e}lian}, {Keitel}, {Key}, {Khadka},
  {Khalili}, {Khan}, {Khan}, {Khazanov}, {Khetan}, {Khursheed}, {Kijbunchoo},
  {Kim}, {Kim}, {Kim}, {Kim}, {Kim}, {Kim}, {Kimball}, {King}, {Kinley-Hanlon},
  {Kirchhoff}, {Kissel}, {Kleybolte}, {Klimenko}, {Knowles}, {Knyazev}, {Koch},
  {Koehlenbeck}, {Koekoek}, {Koley}, {Kolstein}, {Komori}, {Kondrashov},
  {Kontos}, {Koper}, {Korobko}, {Korth}, {Kovalam}, {Kozak}, {Kr{\"a}mer},
  {Kringel}, {Krishnendu}, {Kr{\'o}lak}, {Kuehn}, {Kumar}, {Kumar}, {Kumar},
  {Kumar}, {Kuns}, {Kwang}, {Lackey}, {Laghi}, {Lalande}, {Lam}, {Lamberts},
  {Landry}, {Lane}, {Lang}, {Lange}, {Lantz}, {Lanza}, {La Rosa},
  {Lartaux-Vollard}, {Lasky}, {Laxen}, {Lazzarini}, {Lazzaro}, {Leaci},
  {Leavey}, {Lecoeuche}, {Lee}, {Lee}, {Lee}, {Lee}, {Lehmann}, {Leon},
  {Leroy}, {Letendre}, {Levin}, {Li}, {Li}, {Li}, {Li}, {Li}, {Linde},
  {Linker}, {Linley}, {Littenberg}, {Liu}, {Liu}, {Llorens-Monteagudo}, {Lo},
  {Lockwood}, {London}, {Longo}, {Lorenzini}, {Loriette}, {Lormand}, {Losurdo},
  {Lough}, {Lousto}, {Lovelace}, {L{\"u}ck}, {Lumaca}, {Lundgren}, {Ma},
  {Macas}, {MacInnis}, {Macleod}, {MacMillan}, {Macquet}, {Maga{\~n}a
  Hernandez}, {Maga{\~n}a-Sandoval}, {Magazz{\`u}}, {Magee}, {Majorana},
  {Maksimovic}, {Maliakal}, {Malik}, {Man}, {Mandic}, {Mangano}, {Mansell},
  {Manske}, {Mantovani}, {Mapelli}, {Marchesoni}, {Marion}, {M{\'a}rka},
  {M{\'a}rka}, {Markakis}, {Markosyan}, {Markowitz}, {Maros}, {Marquina},
  {Marsat}, {Martelli}, {Martin}, {Martin}, {Martinez}, {Martinez}, {Martynov},
  {Masalehdan}, {Mason}, {Massera}, {Masserot}, {Massinger}, {Masso-Reid},
  {Mastrogiovanni}, {Matas}, {Mateu-Lucena}, {Matichard}, {Matiushechkina},
  {Mavalvala}, {Maynard}, {McCann}, {McCarthy}, {McClelland}, {McCormick},
  {McCuller}, {McGuire}, {McIsaac}, {McIver}, {McManus}, {McRae}, {McWilliams},
  {Meacher}, {Meadors}, {Mehmet}, {Mehta}, {Melatos}, {Melchor}, {Mendell},
  {Menendez-Vazquez}, {Mercer}, {Mereni}, {Merfeld}, {Merilh}, {Merritt},
  {Merzougui}, {Meshkov}, {Messenger}, {Messick}, {Metzdorff}, {Meyers},
  {Meylahn}, {Mhaske}, {Miani}, {Miao}, {Michaloliakos}, {Michel}, {Middleton},
  {Milano}, {Miller}, {Miller}, {Millhouse}, {Mills}, {Milotti},
  {Milovich-Goff}, {Minazzoli}, {Minenkov}, {Mir}, {Mishkin}, {Mishra},
  {Mistry}, {Mitra}, {Mitrofanov}, {Mitselmakher}, {Mittleman}, {Mo},
  {Mogushi}, {Mohapatra}, {Mohite}, {Molina}, {Molina-Ruiz}, {Mondin},
  {Montani}, {Moore}, {Moraru}, {Morawski}, {Moreno}, {Morisaki}, {Mours},
  {Mow-Lowry}, {Mozzon}, {Muciaccia}, {Mukherjee}, {Mukherjee}, {Mukherjee},
  {Mukherjee}, {Mukund}, {Mullavey}, {Munch}, {Mu{\~n}iz}, {Murray}, {Nadji},
  {Nagar}, {Nardecchia}, {Naticchioni}, {Nayak}, {Neil}, {Neilson}, {Nelemans},
  {Nelson}, {Nery}, {Neunzert}, {Ng}, {Ng}, {Nguyen}, {Nguyen}, {Nguyen},
  {Nichols}, {Nissanke}, {Nocera}, {Noh}, {North}, {Nothard}, {Nuttall},
  {Oberling}, {O'Brien}, {O'Dell}, {Oganesyan}, {Ogin}, {Oh}, {Oh}, {Ohme},
  {Ohta}, {Okada}, {Olivetto}, {Oppermann}, {Oram}, {O'Reilly}, {Ormiston},
  {Ormsby}, {Ortega}, {O'Shaughnessy}, {Ossokine}, {Osthelder}, {Ottaway},
  {Overmier}, {Owen}, {Pace}, {Pagano}, {Page}, {Pagliaroli}, {Pai}, {Pai},
  {Palamos}, {Palashov}, {Palomba}, {Pan}, {Panda}, {Pang}, {Pankow},
  {Pannarale}, {Pant}, {Paoletti}, {Paoli}, {Paolone}, {Parker}, {Pascucci},
  {Pasqualetti}, {Passaquieti}, {Passuello}, {Patel}, {Patricelli}, {Payne},
  {Pechsiri}, {Pedraza}, {Pegoraro}, {Pele}, {Penn}, {Perego}, {Perez},
  {P{\'e}rigois}, {Perreca}, {Perri{\`e}s}, {Petermann}, {Petterson},
  {Pfeiffer}, {Pham}, {Phukon}, {Piccinni}, {Pichot}, {Piendibene},
  {Piergiovanni}, {Pierini}, {Pierro}, {Pillant}, {Pilo}, {Pinard}, {Pinto},
  {Piotrzkowski}, {Pirello}, {Pitkin}, {Placidi}, {Plastino}, {Pluchar},
  {Poggiani}, {Polini}, {Pong}, {Ponrathnam}, {Popolizio}, {Porter},
  {Poverman}, {Powell}, {Pracchia}, {Prajapati}, {Prasai}, {Prasanna},
  {Pratten}, {Prestegard}, {Principe}, {Prodi}, {Prokhorov}, {Prosposito},
  {Puecher}, {Punturo}, {Puosi}, {Puppo}, {P{\"u}rrer}, {Qi}, {Quetschke},
  {Quinonez}, {Quitzow-James}, {Raab}, {Raaijmakers}, {Radkins}, {Radulesco},
  {Raffai}, {Rafferty}, {Rail}, {Raja}, {Rajan}, {Rajbhandari}, {Rakhmanov},
  {Ramirez}, {Ramirez}, {Ramos-Buades}, {Rana}, {Rao}, {Rapagnani}, {Rapol},
  {Ratto}, {Raymond}, {Razzano}, {Read}, {Regimbau}, {Rei}, {Reid}, {Reitze},
  {Rettegno}, {Ricci}, {Richardson}, {Richardson}, {Richardson}, {Ricker},
  {Riemenschneider}, {Riles}, {Rizzo}, {Robertson}, {Robinet}, {Rocchi},
  {Rocha}, {Rodriguez}, {Rodriguez-Soto}, {Rolland}, {Rollins}, {Roma},
  {Romanelli}, {Romano}, {Romel}, {Romero}, {Romero-Shaw}, {Romie}, {Ronchini},
  {Rose}, {Rose}, {Rose}, {Rosell}, {Rosi{\'n}ska}, {Rosofsky}, {Ross},
  {Rowan}, {Rowlinson}, {Roy}, {Roy}, {Ruggi}, {Ryan}, {Sachdev}, {Sadecki},
  {Sadiq}, {Sakellariadou}, {Salafia}, {Salconi}, {Saleem}, {Samajdar},
  {Sanchez}, {Sanchez}, {Sanchez}, {Sanchis-Gual}, {Sanders}, {Santiago},
  {Santos}, {Saravanan}, {Sarin}, {Sassolas}, {Sathyaprakash}, {Sauter},
  {Savage}, {Savant}, {Sawant}, {Sayah}, {Schaetzl}, {Schale}, {Scheel},
  {Scheuer}, {Schindler-Tyka}, {Schmidt}, {Schnabel}, {Schofield},
  {Sch{\"o}nbeck}, {Schreiber}, {Schulte}, {Schutz}, {Schwarm}, {Schwartz},
  {Scott}, {Scott}, {Seglar-Arroyo}, {Seidel}, {Sellers}, {Sengupta},
  {Sennett}, {Sentenac}, {Sequino}, {Sergeev}, {Setyawati}, {Shaffer},
  {Shahriar}, {Sharifi}, {Sharma}, {Sharma}, {Shawhan}, {Shen}, {Shikauchi},
  {Shink}, {Shoemaker}, {Shoemaker}, {Shukla}, {ShyamSundar}, {Sieniawska},
  {Sigg}, {Singer}, {Singh}, {Singh}, {Singha}, {Singhal}, {Sintes}, {Sipala},
  {Skliris}, {Slagmolen}, {Slaven-Blair}, {Smetana}, {Smith}, {Smith},
  {Somala}, {Son}, {Soni}, {Sorazu}, {Sordini}, {Sorrentino}, {Sorrentino},
  {Soulard}, {Souradeep}, {Sowell}, {Spencer}, {Spera}, {Srivastava},
  {Srivastava}, {Staats}, {Stachie}, {Steer}, {Steinke}, {Steinlechner},
  {Steinlechner}, {Steinmeyer}, {Stevenson}, {Stolle-McAllister}, {Stops},
  {Stover}, {Strain}, {Stratta}, {Strunk}, {Sturani}, {Stuver}, {S{\"u}dbeck},
  {Sudhagar}, {Sudhir}, {Suh}, {Summerscales}, {Sun}, {Sun}, {Sunil}, {Sur},
  {Suresh}, {Sutton}, {Swinkels}, {Szczepa{\'n}czyk}, {Tacca}, {Tait},
  {Talbot}, {Tanasijczuk}, {Tanner}, {Tao}, {Tapia}, {Tapia San Martin},
  {Tasson}, {Taylor}, {Tenorio}, {Terkowski}, {Thirugnanasambandam}, {Thomas},
  {Thomas}, {Thomas}, {Thompson}, {Thondapu}, {Thorne}, {Thrane}, {Tiwari},
  {Tiwari}, {Tiwari}, {Toland}, {Tolley}, {Tonelli}, {Tornasi},
  {Torres-Forn{\'e}}, {Torrie}, {Melo}, {T{\"o}yr{\"a}}, {Tran}, {Trapananti},
  {Travasso}, {Traylor}, {Tringali}, {Tripathee}, {Trovato}, {Trudeau}, {Tsai},
  {Tsang}, {Tse}, {Tso}, {Tsukada}, {Tsuna}, {Tsutsui}, {Turconi}, {Ubhi},
  {Udall}, {Ueno}, {Ugolini}, {Unnikrishnan}, {Urban}, {Usman}, {Utina},
  {Vahlbruch}, {Vajente}, {Vajpeyi}, {Valdes}, {Valentini}, {Valsan}, {van
  Bakel}, {van Beuzekom}, {van den Brand}, {Van Den Broeck}, {Vander-Hyde},
  {van der Schaaf}, {van Heijningen}, {Vardaro}, {Vargas}, {Varma}, {Vass},
  {Vas{\'u}th}, {Vecchio}, {Vedovato}, {Veitch}, {Veitch}, {Venkateswara},
  {Venneberg}, {Venugopalan}, {Verkindt}, {Verma}, {Veske}, {Vetrano},
  {Vicer{\'e}}, {Viets}, {Villa-Ortega}, {Vinet}, {Vitale}, {Vo}, {Vocca},
  {Vorvick}, {Vyatchanin}, {Wade}, {Wade}, {Wade}, {Walet}, {Walker},
  {Wallace}, {Wallace}, {Walsh}, {Wang}, {Wang}, {Wang}, {Wang}, {Ward},
  {Warner}, {Was}, {Washington}, {Watchi}, {Weaver}, {Wei}, {Weinert},
  {Weinstein}, {Weiss}, {Wellmann}, {Wen}, {We{\ss}els}, {Westhouse}, {Wette},
  {Whelan}, {White}, {White}, {Whiting}, {Whittle}, {Wilken}, {Williams},
  {Williams}, {Williamson}, {Willis}, {Willke}, {Wilson}, {Wimmer}, {Winkler},
  {Wipf}, {Woan}, {Woehler}, {Wofford}, {Wong}, {Wrangel}, {Wright}, {Wu},
  {Wysocki}, {Xiao}, {Yamamoto}, {Yang}, {Yang}, {Yang}, {Yap}, {Yeeles},
  {Yoon}, {Yu}, {Yu}, {Yuen}, {Zadro{\.z}ny}, {Zanolin}, {Zelenova}, {Zendri},
  {Zevin}, {Zhang}, {Zhang}, {Zhang}, {Zhang}, {Zhao}, {Zhao}, {Zhou}, {Zhou},
  {Zhu}, {Zimmerman}, {Zucker}, \& {Zweizig}}]{2020arXiv201014533T}
{The LIGO Scientific Collaboration}, {the Virgo Collaboration}, {Abbott}, R.,
  {et~al.} 2020, arXiv e-prints, arXiv:2010.14533.
\newblock \doarXiv{2010.14533}

\bibitem[{{Thompson} {et~al.}(2004){Thompson}, {Chang}, \&
  {Quataert}}]{Thompson2004}
{Thompson}, T.~A., {Chang}, P., \& {Quataert}, E. 2004, \apj, 611, 380,
  \dodoi{10.1086/421969}

\bibitem[{{Torrealba} {et~al.}(2018){Torrealba}, {Belokurov}, {Koposov},
  {Bechtol}, {Drlica-Wagner}, {Olsen}, {Vivas}, {Yanny}, {Jethwa}, {Walker},
  {Li}, {Allam}, {Conn}, {Gallart}, {Gruendl}, {James}, {Johnson}, {Kuehn},
  {Kuropatkin}, {Martin}, {Martinez-Delgado}, {Nidever}, {No{\"e}l}, {Simon},
  {Stringfellow}, \& {Tucker}}]{Torrealba18}
{Torrealba}, G., {Belokurov}, V., {Koposov}, S.~E., {et~al.} 2018, \mnras, 475,
  5085, \dodoi{10.1093/mnras/sty170}

\bibitem[{{Tsujimoto} \& {Shigeyama}(2014)}]{Tsujimoto+Shigeyama14}
{Tsujimoto}, T., \& {Shigeyama}, T. 2014, \aap, 565, L5,
  \dodoi{10.1051/0004-6361/201423751}

\bibitem[{{van de Voort} {et~al.}(2015){van de Voort}, {Quataert}, {Hopkins},
  {Kere{\v s}}, \& {Faucher-Gigu{\`e}re}}]{vandeVoort15}
{van de Voort}, F., {Quataert}, E., {Hopkins}, P.~F., {Kere{\v s}}, D., \&
  {Faucher-Gigu{\`e}re}, C.-A. 2015, \mnras, 447, 140,
  \dodoi{10.1093/mnras/stu2404}

\bibitem[{{Venn} {et~al.}(2004){Venn}, {Irwin}, {Shetrone}, {Tout}, {Hill}, \&
  {Tolstoy}}]{Venn04}
{Venn}, K.~A., {Irwin}, M., {Shetrone}, M.~D., {et~al.} 2004, \aj, 128, 1177,
  \dodoi{10.1086/422734}

\bibitem[{{Wanderman} \& {Piran}(2010)}]{Wanderman2010}
{Wanderman}, D., \& {Piran}, T. 2010, \mnras, 406, 1944,
  \dodoi{10.1111/j.1365-2966.2010.16787.x}

\bibitem[{{Wang} {et~al.}(2020){Wang}, {Zou}, {Liu}, {Liao}, {Liu}, {Chai}, \&
  {Xia}}]{LGRB_stats}
{Wang}, F., {Zou}, Y.-C., {Liu}, F., {et~al.} 2020, \apj, 893, 77,
  \dodoi{10.3847/1538-4357/ab0a86}

\bibitem[{{Wardle} \& {Knapp}(1986)}]{wardle86}
{Wardle}, M., \& {Knapp}, G.~R. 1986, \aj, 91, 23, \dodoi{10.1086/113976}

\bibitem[{{Wehmeyer} {et~al.}(2015){Wehmeyer}, {Pignatari}, \&
  {Thielemann}}]{Wehmeyer15}
{Wehmeyer}, B., {Pignatari}, M., \& {Thielemann}, F.~K. 2015, \mnras, 452,
  1970, \dodoi{10.1093/mnras/stv1352}

\bibitem[{{Wong} {et~al.}(2010){Wong}, {Willems}, \& {Kalogera}}]{Wong2010}
{Wong}, T.-W., {Willems}, B., \& {Kalogera}, V. 2010, \apj, 721, 1689,
  \dodoi{10.1088/0004-637X/721/2/1689}

\bibitem[{{Zevin} {et~al.}(2019){Zevin}, {Kremer}, {Siegel}, {Coughlin},
  {Tsang}, {Berry}, \& {Kalogera}}]{Zevin19}
{Zevin}, M., {Kremer}, K., {Siegel}, D.~M., {et~al.} 2019, \apj, 886, 4,
  \dodoi{10.3847/1538-4357/ab498b}

\end{thebibliography}



\appendix

\section{Average Iron Yield from Core-collapse Supernovae} \label{subsec:yFe}

To estimate an average iron yield from core-collapse supernovae (CCSN), we calculate a weighted average between observations of H-rich CCSN and H-poor CCSN. The majority of the iron comes from the Ni-56 $\rightarrow$ Co-56 $\rightarrow$ Fe-56 decay chain, so we adopt mean Ni-56 yields for both Type II SN and Type IIb/Ib/Ic SN as measured from modeling their bolometric light curves, and account for the relative rate between these two broad classes. The mean Ni-56 yield from H-rich CCSN is $0.044 \pm 0.044 M_\odot$ \citep{Anderson19}. The mean Ni-56 yield from Type II CCSN is $0.12 \pm 0.12 M_\odot$ \citep{Afsariardchi2020}. For the relative rates of stripped envelope to Type II CCSN, we adopt the results of the Lick Observatory SN Search (LOSS; \citealt{LOSS11}). While LOSS was a targeted survey, it remains the most complete volume-limited supernova search completed to date.  We consider two cases: (1) Based on the entire LOSS sample, \citet{2017PASP..129e4201S} find the relative fractions of Type II and stripped envelope SN of $69.6^{+6.7}_{-6.7}$\% and $30.4^{+5.0}_{-4.9}$\%, respectively. This implies an average iron yield of $0.067 M_\odot$. (2) However, the ratio of Type II to stripped envelope SN is metallicity dependent. \citet{Graur2017a} examined the relative rates the LOSS sample as a function of host galaxy mass and metallicity. In their lowest metallicity bin, the ratio of SESN and Type II SN specific rates is $R_{SE}/R_{II} = 0.13^{+0.09}_{-0.08}$, a factor of three lower than the overall LOSS sample above. Adopting this value, and making the (rather large) assumption that the the average nickel yield of each class is not a function of metallicity, we find an average CCSN iron yield of $0.053 M_\odot$. In both (1) and (2), the average yield is slightly below the order of $y_{\text{Fe}} \approx 0.1 M_\odot$, with the uncertainty of $f_{retained}$ and $M_{gas}$ far outweighing that of $y_{\text{Fe}}$.

\end{document}